\documentclass[twocolumn,aps,pra,amssymb,amstext,showpacs]{revtex4-1}

\usepackage{amsfonts,amssymb,amsmath,graphicx,color}
\pdfoutput=1

\newcommand{\real}{\mathrm{Re}}

\begin{document}

\title{Manipulating quantum wave packets via time-dependent
  absorption}

\author{Arseni Goussev}

\affiliation{Department of Mathematics and Information Sciences,
  Northumbria University, Newcastle Upon Tyne, NE1 8ST, United
  Kingdom}

\date{\today}

\begin{abstract}
  A pulse of matter waves may dramatically change its shape when
  traversing an absorbing barrier with time-dependent
  transparency. Here we show that this effect can be utilized for
  controlled manipulation of spatially-localized quantum states. In
  particular, in the context of atom-optics experiments, we explicitly
  demonstrate how the proposed approach can be used to generate
  spatially shifted, split, squeezed and cooled atomic wave
  packets. We expect our work to be useful in devising new
  interference experiments with atoms and molecules and, more
  generally, to enable new ways of coherent control of matter waves.
\end{abstract}


\maketitle


\section{Introduction}
\label{sec:intro}

The ability to engineer and manipulate the spatial wave function of a
quantum particle has far-reaching applications in many areas of
physics. One example is the so-called ``beam-splitting'', a coherent
division of an atomic or molecular wave function into two or more
non-overlapping wave packets (WPs), which is indispensable in
matter-wave interferometry
\cite{CSP09Optics,*HGH+12Colloquium,*Arn14DeBroglie}.  Another example
is generation of ``squeezed states'', i.e., WPs with a reduced
uncertainty in one dynamical variable (at the expense of an increased
uncertainty in the conjugate variable) that are widely used to enhance
the precision of quantum measurements \cite{GLM04Quantum}. To date,
various mechanisms of reshaping matter WPs have been explored, some of
which utilize time-dependent harmonic traps \cite{CD96Bose}, quantum
holography \cite{WAB99Controlling}, spatially homogeneous laser light
with time-dependent amplitude in the presence of a spatially
inhomogeneous magnetic field \cite{ODHP00deBroglie}, periodic
potentials \cite{ETA+03Dispersion}, time-dependent electric fields in
atom chips \cite{NJF+10Dynamic}, finite-length attractive optical
lattices with a slowly varying envelope
\cite{FCG+11Realization,*CFV+13Matter}, or chaotic scattering dynamics
\cite{GCR+12Optically}. Here we report an alternative approach based
on the principle of time-dependent absorption that has versatile
applications.

\begin{figure}[h]
\includegraphics[width=2.8in]{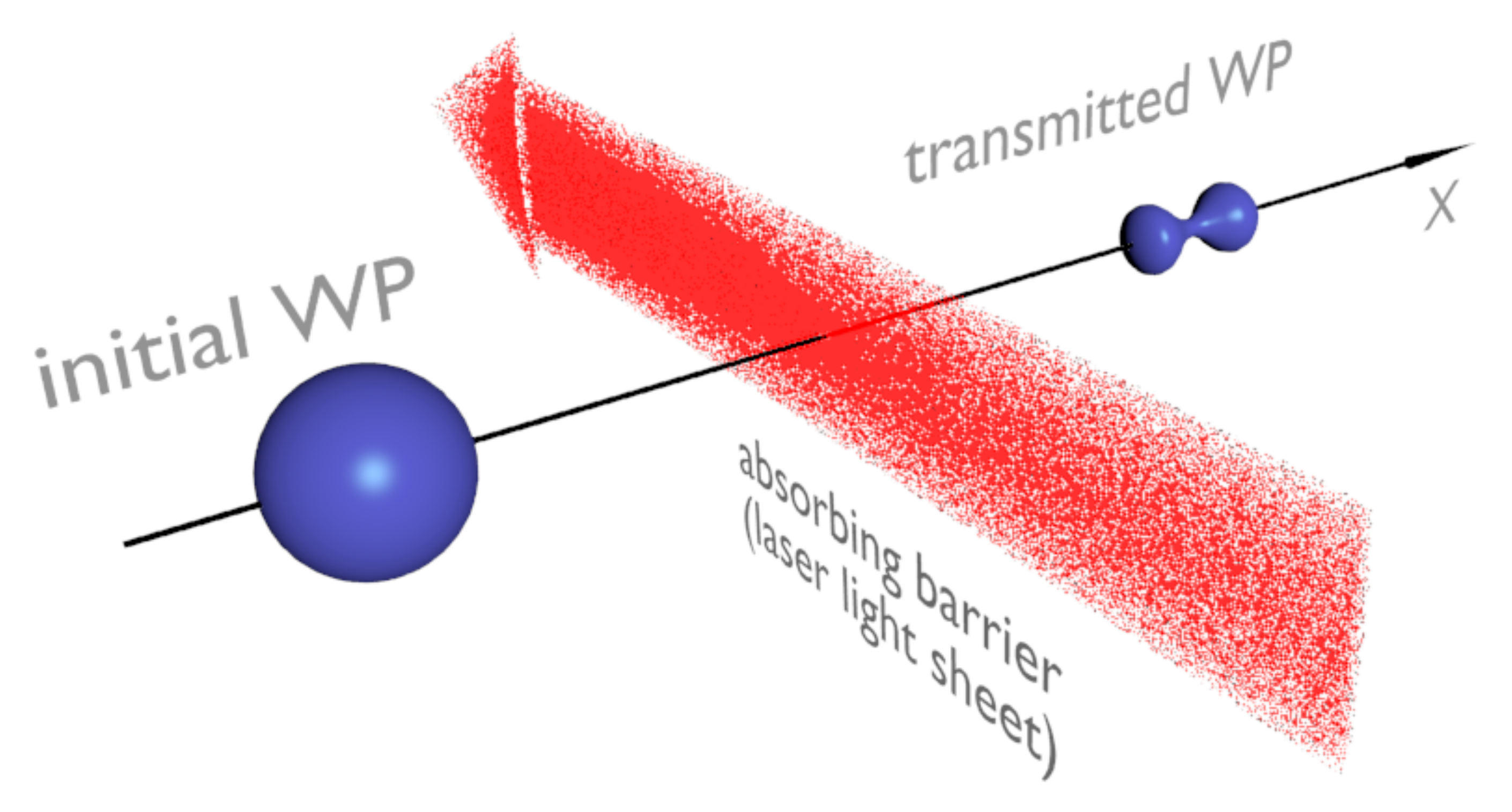}
\caption{Illustration of matter pulse carving.}
\label{fig1}
\end{figure}

In this paper, we show that a localized quantum WP can be efficiently
manipulated -- spatially shifted, split, squeezed and cooled -- by
making it pass through a time-dependent absorbing barrier, i.e., a
narrow region of space acting as a particle density sink (see
Ref.~\cite{MPNE04Complex} for mechanisms and treatment of absorption
in quantum systems). The absorbing barrier can be realized in the
context of atom-optics with a sheet of laser light intercepting the
motion of an atomic cloud (see Fig.~\ref{fig1}). The radiation
frequency of the laser can be chosen such that a passing atom becomes
undetectable due to ionization or a change of its internal state, and
thus effectively absorbed. Furthermore, the strength of this
absorption process can be made time-dependent by varying the intensity
of the laser in accordance with an externally prescribed function of
time.

To further exemplify the process of time-dependent absorption, we
consider an atom initially (at $t = 0$) prepared in a state
$\Psi^{(0)}(x) | \mathrm{in} \rangle$, where $| \mathrm{in}\rangle$
denotes an internal state of the atom, and $\Psi^{(0)}(x)$ is a
spatially localized WP representing its center-of-mass motion. As the
atom traverses a laser light sheet positioned at $x=0$, the laser may
trigger an atomic transition from $| \mathrm{in} \rangle$ to another
state (or to one of several other states) $| \mathrm{abs} \rangle$ of
the atomic spectrum. The probability of the transition is nonzero only
inside the light sheet, i.e,. in a close vicinity of the point $x=0$,
and can be made to depend on time by externally modifying the
intensity of the laser. At some time $t > 0$, the atom will be found
in a state $\Psi^{(t)}(x) | \mathrm{in} \rangle +
\sum\limits_{\mathrm{abs}} \psi_{\mathrm{abs}}^{(t)}(x) | \mathrm{abs}
\rangle$, where $\Psi^{(t)}(x)$ and $\psi_{\mathrm{abs}}^{(t)}(x)$
denote spatial wave functions corresponding to the internal states $|
\mathrm{in} \rangle$ and $| \mathrm{abs} \rangle$, respectively. In
what follows below, we will only be concerned with a projection of the
full atomic state on $| \mathrm{in} \rangle$, and will regard the rest
of the state, $\sum\limits_{\mathrm{abs}} \psi_{\mathrm{abs}}^{(t)}(x)
| \mathrm{abs} \rangle$, as a part that has been removed, or
``absorbed'', by the light sheet. It is in this sense that we will be
interested in a transformation of the center-of-mass wave function
$\Psi$, from $\Psi^{(0)}(x)$ to $\Psi^{(t)}(x)$, induced by the light
sheet playing the role of an absorbing barrier.

The problem of a non-relativistic quantum particle interacting with an
absorbing barrier has a long history and is a paradigm of the theory
of quantum transients \cite{Kle94Exact,*CGM09Quantum}. The limit of an
instantaneously opening or closing barrier, first addressed by
Moshinsky, was shown to cause well-pronounced oscillations of the
probability density distribution; this effect is known as
``diffraction in time'' (DIT) and bears close mathematical analogy to
optical diffraction of light at an aperture with straight edges
\cite{Mos52Diffraction,*GK76Possibility,*Mos76Diffraction,*BZ97Diffraction,*MMS99Diffraction,*God02Diffraction,*God03Diffraction,*GM05Generic,*TMB+11Explanation}. DIT
is generally suppressed if the barrier switching is not sufficiently
abrupt \cite{CMM07Time} or if the particle momentum distribution
exhibits incoherent thermal broadening \cite{God09Statistical}. In a
many particle scenario, the interaction between particles is also
known to suppress DIT \cite{CM06Dynamics}.

Here we address the motion of a quantum WP in the presence of an
absorbing barrier that may, during some intervals of time, open or
close exponentially fast, but yet not fast enough to trigger DIT. We
show that this subtle regime, being well within experimental reach,
allows for controlled reshaping of the WP through the process of
removing (or ``carving out'') parts of the probability density profile
without a side effect of producing diffraction ripples. Our analysis
takes into account finite temperature effects, which are inevitable in
any laboratory experiment.

The paper is organized as follows. In Sec.~\ref{sec:theory}, we
present a general framework for an analytical description of the
motion of a quantum particle in the presence of a time-dependent
absorbing barrier. In Sec.~\ref{sec:engineering}, we demonstrate the
possibility of controlled manipulation of a spatial wave function of
the particle, and, in particular, show how the wave function can be
displaced (Sec.~\ref{sec:shift}), split (Sec.~\ref{sec:split}), and
squeezed (Sec.~\ref{sec:squeeze}). We make concluding remarks in
Sec.~\ref{sec:conclusion}. Technical details are deferred to two
Appendixes.


\section{Theoretical framework}
\label{sec:theory}

In order to facilitate analytical treatment, we consider a quantum
particle of mass $m$ initially represented by a Gaussian WP
\begin{align}
  &\Psi^{(0)}(x) = \psi_{\alpha_0, x_0, v_0}^{(0)}(x)
  \nonumber\\ &\quad \equiv \left( \frac{2 \real (\alpha_0)}{\pi}
  \right)^{1/4} e^{ -\alpha_0 (x - x_0)^2 + i m v_0 (x - x_0) / \hbar
  } \,.
\end{align}
Here, $x_0$ and $v_0$ represent respectively the mean position and
velocity of the particle, and the parameter $\alpha_0$ is related to
the spatial extent $\sigma$ of the WP through $\sigma^{-2} = 2 \real
(\alpha_0)$. Hereinafter however we assume $\alpha_0$ to be strictly
positive, so that $\alpha_0 = (2 \sigma^2)^{-1}$. Also, for
concreteness, we take $x_0 < 0$, $\sigma \ll |x_0|$, and $v_0 > 0$. In
other words, the initial WP is assumed to be localized on the
semi-infinite interval $x < 0$ and moving towards the origin.

After a time $t$ and in the absence of any external forcing, the state
$\psi_{\alpha_0, x_0, v_0}^{(0)}$ would evolve into
\begin{align}
  \psi_{\alpha_0, x_0, v_0}^{(t)}(x) &= \int_{-\infty}^{\infty} dx' \,
  K_0^{(t)}(x-x') \psi_{\alpha_0, x_0, v_0}^{(0)}(x') \nonumber \\ &=
  e^{i \phi_t} \psi_{\alpha_t, x_t, v_0}^{(0)}(x) \,,
\label{psi_free_flight}
\end{align}
where
\begin{equation}
  K_0^{(\tau)}(\xi) = \sqrt{\frac{m}{2 \pi i \hbar \tau}} \exp \left(
  i \frac{m \xi^2}{2 \hbar \tau} \right)
\label{free_particle_propagator}
\end{equation}
is the free-particle propagator, and the functions $\alpha_{\tau}$,
$x_{\tau}$, and $\phi_{\tau}$ are defined as
\begin{align}
  &\alpha_{\tau} = \frac{\alpha_0}{1 + 2 i \hbar \alpha_0 \tau / m}
  \,, \\ \quad &x_{\tau} = x_0 + v_0 \tau \,, \\ \quad &\phi_{\tau} =
  \frac{m v_0^2 \tau}{2 \hbar} - \frac{1}{2} \tan^{-1} \left( \frac{2
    \hbar \alpha_0}{m} \tau \right) \,.
\end{align}

We further imagine that in the course of its motion the particle
encounters an infinitesimally thin absorbing barrier positioned at $x
= 0$. (In a realistic setting, the width of the barrier is assumed to
be much smaller than the WP size.) The time-dependent transparency of
the barrier is characterized by a real-valued aperture function
$\chi_{\tau}$, ranging between 0 (representing zero
transparency/complete absorption) and 1 (complete transparency/zero
absorption). A propagator, transporting the particle probability
amplitude from a source point $x' < 0$ to a point $x > 0$ on the other
side of the barrier in time $t > 0$, can be written as
\cite{Gou12Huygens,Gou13Diffraction}
\begin{equation}
  K^{(t)}(x,x') = \int\limits_0^t d\tau \, \frac{\chi_{\tau}}{2}
  \left( \frac{x}{t-\tau} - \frac{x'}{\tau} \right) K_0^{(t-\tau)}(x)
  K_0^{(\tau)}(x') \,.
\label{main_propagator}
\end{equation}
The structure of the propagator $K^{(t)}(x,x')$ stems from a
superposition of a continuous family of paths, parametrized by time
$\tau$. The propagation of the particle along each of these paths
consists of three consecutive stages: First, the particle moves freely
from the source point $x' < 0$ to the barrier at the origin in time
$\tau$; second, the probability amplitude gets modulated by a factor
proportional to the transparency of the barrier $\chi_{\tau}$; third,
the particle travels freely to the observation point $x > 0$ through
the remaining time $(t-\tau)$. The additional factor $\frac{1}{2}
\left( \frac{x}{t-\tau} - \frac{x'}{\tau} \right)$ has the meaning of
the average classical velocity at which the particle crosses the
barrier, and is essential for correctly weighing relative
contributions of the interfering paths.

In fact, the propagator $K^{(t)}(x,x')$, given for $x>0$ by
Eq.~(\ref{main_propagator}), is an exact solution of the following
quantum-mechanical problem (see Ref.~\cite{Gou13Diffraction} for full
details). In this model $K^{(t)}(x,x')$, transporting a wave function
from a source point $x'<0$ to a point $x \not= 0$, is set to satisfy
(i) the time-dependent Schr\"{o}dinger equation
\begin{equation}
  \left( i\hbar \frac{\partial}{\partial \tau} + \frac{\hbar^2}{2 m}
  \frac{\partial^2}{\partial x^2} \right) K^{(\tau)}(x,x') = 0
\end{equation}
for $0 < \tau < t$ and both $x<0$ and $x>0$, (ii) the initial
condition $K^{(0^+)}(x,x') = \delta(x-x')$, (iii) Dirichlet boundary
conditions at $x=\pm \infty$ for negative imaginary times, i.e,
$K^{(-i|\tau|)}(\pm \infty, x') = 0$, and (iv) two matching conditions
relating the propagator and its spatial derivative at $x<0$ to those
at $x>0$, namely
\begin{align}
  K^{(\tau)}(x,x') \big|_{x=0^-}^{x=0^+} &= -(1 - \chi_{\tau})
  K_0^{(\tau)}(x-x') \big|_{x=0} \,, \label{match_cond_1} \\[0.2cm]
  \frac{\partial K^{(\tau)}(x,x')}{\partial x} \bigg|_{x=0^-}^{x=0^+}
  &= -(1 - \chi_{\tau}) \frac{\partial K_0^{(\tau)}(x-x')}{\partial x}
  \bigg|_{x=0} \,, \label{match_cond_2}
\end{align}
for $0 < \tau < t$. Here, $K_0$ denotes the free-particle propagator
defined by Eq.~(\ref{free_particle_propagator}). The matching
conditions (\ref{match_cond_1}) and (\ref{match_cond_2}) are a
time-dependent quantum-mechanical version of the absorbing
(``black-screen'') boundary conditions proposed by Kottler in the
context of stationary wave optics \cite{Kot23Zur,*Kot65Diffraction};
in their original time-independent formulation, Kottler boundary
conditions can be viewed as a mathematical justification of Kirchhoff
diffraction theory.

The wave function transmitted through the barrier at time $t$ is given
by
\begin{align}
  &\Psi^{(t)}(x) = \Psi_{\alpha_0, x_0, v_0}^{(t)}(x) \equiv
  \int\limits_{-\infty}^{\infty} dx' \, K^{(t)}(x,x') \psi_{\alpha_0,
    x_0, v_0}^{(0)} (x') \nonumber \\ &\; = \int\limits_0^t d\tau \,
  \frac{\chi_{\tau}}{2} \left( \frac{x}{t-\tau} + \frac{\alpha_{\tau}
    v_0}{\alpha_{t_0}} \right) K_0^{(t-\tau)}(x) \psi_{\alpha_0, x_0,
    v_0}^{(\tau)} (0) \,,
\label{Psi(x,t)}
\end{align}
where $t_0 = |x_0| / v_0$ represents the time needed for the
corresponding classical particle to reach the barrier. Here however we
choose to focus on a phase-space representation of the wave function
as provided by the Husimi distribution \cite{[{See, e.g.,
  }][{}]Bal14Quantum}
\begin{equation}
  H_{\alpha_0, x_0, v_0}^{(t)}(\tilde{x}, \tilde{v}) = \left| \langle
  \psi_{\alpha_0, \tilde{x}, \tilde{v}}^{(0)} | \Psi_{\alpha_0, x_0,
    v_0}^{(t)} \rangle \right|^2 \,.
\label{HUSIMI}
\end{equation}
The Husimi distribution quantifies the overlap between the
time-evolved state $\Psi_{\alpha_0, x_0, v_0}^{(t)}$ and a probe
Gaussian WP centered in phase space around $(\tilde{x},\tilde{v})$ and
characterized by the spatial dispersion $\sigma = 1 / \sqrt{2
  \alpha_0}$. Hereinafter, we assume $0 < \sigma \ll \tilde{x}$; this
implies that we only examine the wave function $\Psi_{\alpha_0, x_0,
  v_0}^{(t)}(x)$ deep inside the transmission region. Then, using
Eq.~(\ref{Psi(x,t)}) we obtain
\begin{align}
  \langle \psi_{\alpha_0, \tilde{x}, \tilde{v}}^{(0)} |
  &\Psi_{\alpha_0, x_0, v_0}^{(t)} \rangle = \int\limits_0^t d\tau \,
  \frac{\chi_{\tau}}{2} \left( \frac{\alpha_{t-\tau}
    \tilde{v}}{\alpha_{\tilde{t}}} + \frac{\alpha_{\tau}
    v_0}{\alpha_{t_0}} \right) \nonumber \\ &\qquad \times \left[
    \psi_{\alpha_0, \tilde{x}, \tilde{v}}^{(\tau-t)} (0) \right]^* \,
  \psi_{\alpha_0, x_0, v_0}^{(\tau)} (0) \,,
\label{husimi}
\end{align}
where $\tilde{t} = \tilde{x} / \tilde{v}$ and the asterisk represents
complex conjugation. The integrand in Eq.~(\ref{husimi}), unlike that
in Eq.~(\ref{Psi(x,t)}), is free of singular points on the closed
interval $0 \leq \tau \leq t$. This fact makes formula (\ref{husimi})
an efficient tool for analytical and numerical inspection of the part
of the WP transmitted through the absorbing barrier.

Non-zero temperature, ubiquitous in any laboratory experiment,
manifests itself as an incoherent broadening of the initial velocity
distribution of the particle. In order to account for this broadening,
we describe the particle state in terms of a time-dependent density
matrix
\begin{align}
  &\rho_{\alpha_0, x_0, v_0, \Delta v}^{(\tau)} \nonumber\\ &\quad =
  \frac{1}{\sqrt{\pi}} \int_{-\infty}^{\infty} \frac{dv}{\Delta v} \,
  e^{-(v - v_0)^2 / \Delta v^2} | \Psi_{\alpha_0, x_0, v}^{(\tau)}
  \rangle \langle \Psi_{\alpha_0, x_0, v}^{(\tau)} | \,,
\end{align}
where $\Delta v$ quantifies the width of the thermal spread of the
initial velocity. The corresponding finite-temperature Husimi
distribution reads
\begin{align}
  &\mathcal{H}_{\alpha_0, x_0, v_0, \Delta v}^{(t)}(\tilde{x},
  \tilde{v}) = \mathrm{tr} \left( \rho_{\alpha_0, x_0, v_0, \Delta
    v}^{(t)} | \psi_{\alpha_0, \tilde{x}, \tilde{v}}^{(0)} \rangle
  \langle \psi_{\alpha_0, \tilde{x}, \tilde{v}}^{(0)} | \right)
  \nonumber \\ &\qquad = \frac{1}{\sqrt{\pi}}
  \int\limits_{-\infty}^{\infty} \frac{dv}{\Delta v} \, e^{-(v -
    v_0)^2 / \Delta v^2} H_{\alpha_0, x_0, v}^{(t)}(\tilde{x},
  \tilde{v}) \,,
\label{HUSIMI_temperature}
\end{align}
where tr denotes the trace. Substituting Eqs.~(\ref{HUSIMI}) and
(\ref{husimi}) into Eq.~(\ref{HUSIMI_temperature}), we obtain
\begin{widetext}
\begin{equation}
  \mathcal{H}_{\alpha_0, x_0, v_0, \Delta v}^{(t)}(\tilde{x},
  \tilde{v}) = \frac{1}{4} \int\limits_0^t d\tau \int\limits_0^t
  d\tau' \, \chi_{\tau} \chi_{\tau'} \left[ \psi_{\alpha_0, \tilde{x},
      \tilde{v}}^{(\tau-t)}(0) \right]^* \psi_{\alpha_0, \tilde{x},
    \tilde{v}}^{(\tau'-t)}(0) \; \Phi_{\alpha_0, \tilde{x}, \tilde{v},
    x_0, v_0, \Delta v}^{(t,\tau,\tau')} \,,
\label{HUSIMI_temperature_2}
\end{equation}
where
\begin{equation}
  \Phi_{\alpha_0, \tilde{x}, \tilde{v}, x_0, v_0, \Delta
    v}^{(t,\tau,\tau')} = \frac{1}{\sqrt{\pi}}
  \int\limits_{-\infty}^{\infty} \frac{dv}{\Delta v} \, e^{-(v -
    v_0)^2 / \Delta v^2} \left( \frac{\alpha_{t-\tau}
    \tilde{v}}{\alpha_{\tilde{t}}} + \frac{\alpha_{\tau}
    v}{\alpha_{|x_0|/v}} \right) \left( \frac{\alpha_{t-\tau'}
    \tilde{v}}{\alpha_{\tilde{t}}} + \frac{\alpha_{\tau'}
    v}{\alpha_{|x_0|/v}} \right)^{\!*} \psi_{\alpha_0, x_0, v}^{(\tau)}(0)
  \left[ \psi_{\alpha_0, x_0, v}^{(\tau')}(0) \right]^* .
\label{Phi_function}
\end{equation}
\end{widetext}
In view of the identity $\frac{v}{\alpha_{|x_0|/v}} =
\frac{v}{\alpha_0} + \frac{2 i \hbar |x_0|}{m}$, the integral in
Eq.~(\ref{Phi_function}) is Gaussian in nature and can be evaluated
analytically. (See Appendix~\ref{app1} for the calculation and exact
expression.)

Below we use the Husimi distributions given by Eqs.~(\ref{HUSIMI}) and
(\ref{husimi}) and Eqs.~(\ref{HUSIMI_temperature_2}) and
(\ref{Phi_function}) to analyze several scenarios of WP engineering
corresponding to specific forms of the aperture function
$\chi_{\tau}$.


\section{Wave packet engineering}
\label{sec:engineering}

\subsection{Spatial shifting}
\label{sec:shift}

We first investigate physical effects produced by an exponentially
opening (closing) barrier. Thus, we consider
\begin{equation}
  \chi_{\tau} = \chi_0 \, e^{\gamma \tau} \,,
\label{chi_exp}
\end{equation}
where $\gamma$ is the rate of change of the barrier transparency and
$\chi_0$ is its initial value. While an explicit evaluation of the
corresponding Husimi distribution is not feasible, a significant
analytical insight can be gained in an asymptotic regime, defined by
\begin{equation}
  1 \ll \frac{|x_0|}{\sigma} \lesssim \frac{v_0 t}{2 \sigma} \ll
  \frac{\sigma}{2 \lambdabar} \,,
\label{main_assumption}
\end{equation}
where $\lambdabar = \hbar / (m v_0)$ is the reduced de Broglie
wavelength of the particle. The first of the three conditions combined
in Eq.~(\ref{main_assumption}), $\sigma \ll |x_0|$, has already been
used in deriving Eq.~(\ref{Psi(x,t)}). The second inequality, $2 |x_0|
\lesssim v_0 t$, is needed to ensure that, at time $t$, the particle
has ``passed'' the barrier and the dominant part of the WP is well
localized in the transmission region; under this condition, the Husimi
distribution of the transmitted WP is accurately represented by
Eqs.~(\ref{HUSIMI}) and (\ref{husimi}). Finally, the third inequality,
$\lambdabar \ll \sigma^2 / (v_0 t)$, expresses the semiclassical
limit, in which time variations of the spatial width of the WP can be
effectively neglected; this limit is commonly known as a ``frozen
Gaussian'' regime \cite{Hel81Frozen}.

Adopting the semiclassical regime specified by
Eq.~(\ref{main_assumption}) and further assuming that
\begin{equation}
  |\gamma| \ll \frac{2 |x_0| v_0}{\sigma^2} \,,
\label{restriction_on_gamma}
\end{equation}
we use the method of steepest descent to evaluate the pure-state
Husimi distribution $H_{\alpha_0, x_0, v_0}^{(t)}(\tilde{x},
\tilde{v})$ analytically. (See Appendix~\ref{app2} for details of the
calculation.) In particular, we show that $H_{\alpha_0, x_0,
  v_0}^{(t)}(\tilde{x}, \tilde{v})$ is peaked at the phase-space point
$(\tilde{x}, \tilde{v}) = (x_t + \Delta x, v_0)$ with
\begin{equation}
  \Delta x = - \frac{\gamma \sigma^2}{v_0} \,.
\label{shift}
\end{equation}
This means that the transmitted WP appears to be spatially shifted
compared to the counterpart free-particle WP centered at $(x_t,
v_0)$. The shift is proportional to the rate of change of the barrier
transparency and can be positive (advanced WP) as well as negative
(delayed WP). The average velocity of the particle however
remains unaffected by the barrier. The analysis also reveals that the
overall transmission probability is given approximately by
$\chi_{t_0}^2$.

In order to further explore the effect of the WP shift, we compute the
pure-state and finite-temperature Husimi distributions by numerically
evaluating the integrals in Eqs.~(\ref{husimi}) and
(\ref{HUSIMI_temperature_2}) respectively. To this end, we consider a
cloud of ultra-cold atoms characterized by $m = 86.909$~u (the mass of
a $^{87}$Rb atom), $\sigma = 30$~$\mu$m, $v_0 = 3$~mm/s, and $\Delta v
= 0.1$~mm/s. (A cloud of magnetically levitated $^{87}$Rb atoms with
similar parameters has been recently used by Jendrzejewski et al.~to
experimentally demonstrate coherent backscattering of ultra-cold atoms
in a disordered potential \cite{JMR+12Coherent}.) The cloud is
initially centered at $x_0 = -0.15$~mm and the propagation time is set
to be $t = 100$~ms, implying that $x_t = 0.15$~mm $= |x_0|$ and $t_0 =
50$~ms $= t/2$. (In the experiment in Ref.~\cite{JMR+12Coherent}, it
was possible to let the atomic cloud evolve for as long as 150~ms
before performing imaging.)  Since for the chosen set of parameters
$\lambdabar \simeq 244$ nm, the system is in the semiclassical regime
specified by Eq.~(\ref{main_assumption}). Furthermore, the value of
$|\gamma|$ in all computations below is taken not to exceed
225~s$^{-1}$, which ensures that the restriction given by
Eq.~(\ref{restriction_on_gamma}) is fulfilled.

\begin{figure}[h]
\includegraphics[width=3.4in]{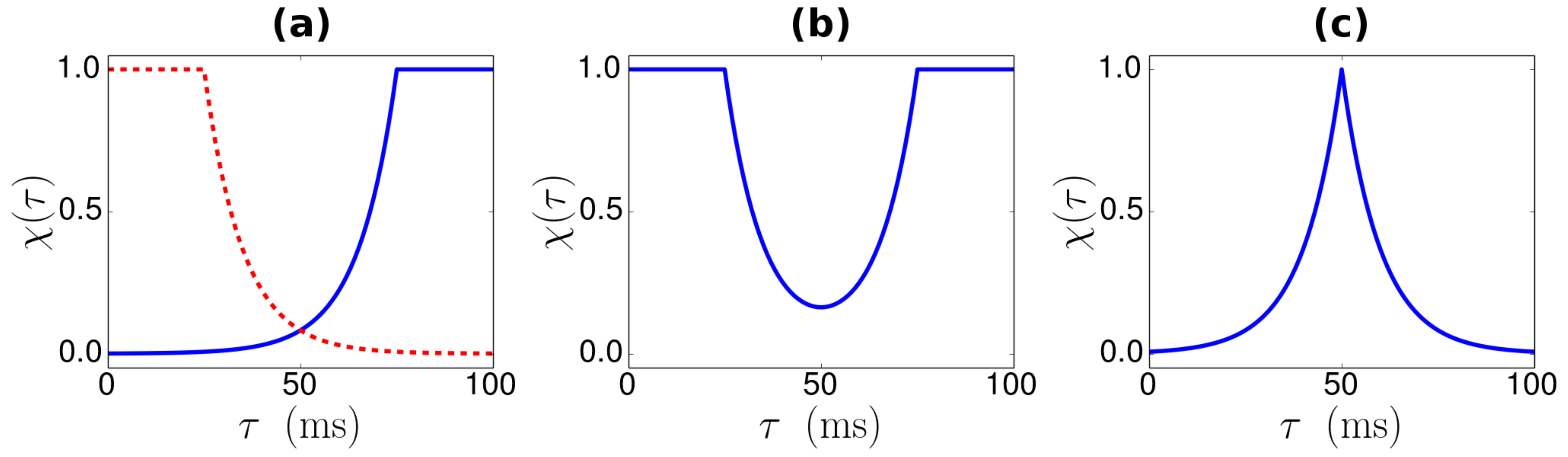}
\caption{Aperture function given by (a) Eq.~(\ref{chi-shift}) with
  $\gamma = 100$~s$^{-1}$ (solid blue curve) and $\gamma =
  -100$~s$^{-1}$ (dashed red curve), (b) Eq.~(\ref{chi-split}) with
  $\gamma = 100$~s$^{-1}$, and (c) Eq.~(\ref{chi-squeeze}) with
  $\gamma = 100$~s$^{-1}$.}
\label{fig2}
\end{figure}

\begin{figure}[h]
\includegraphics[width=3.2in]{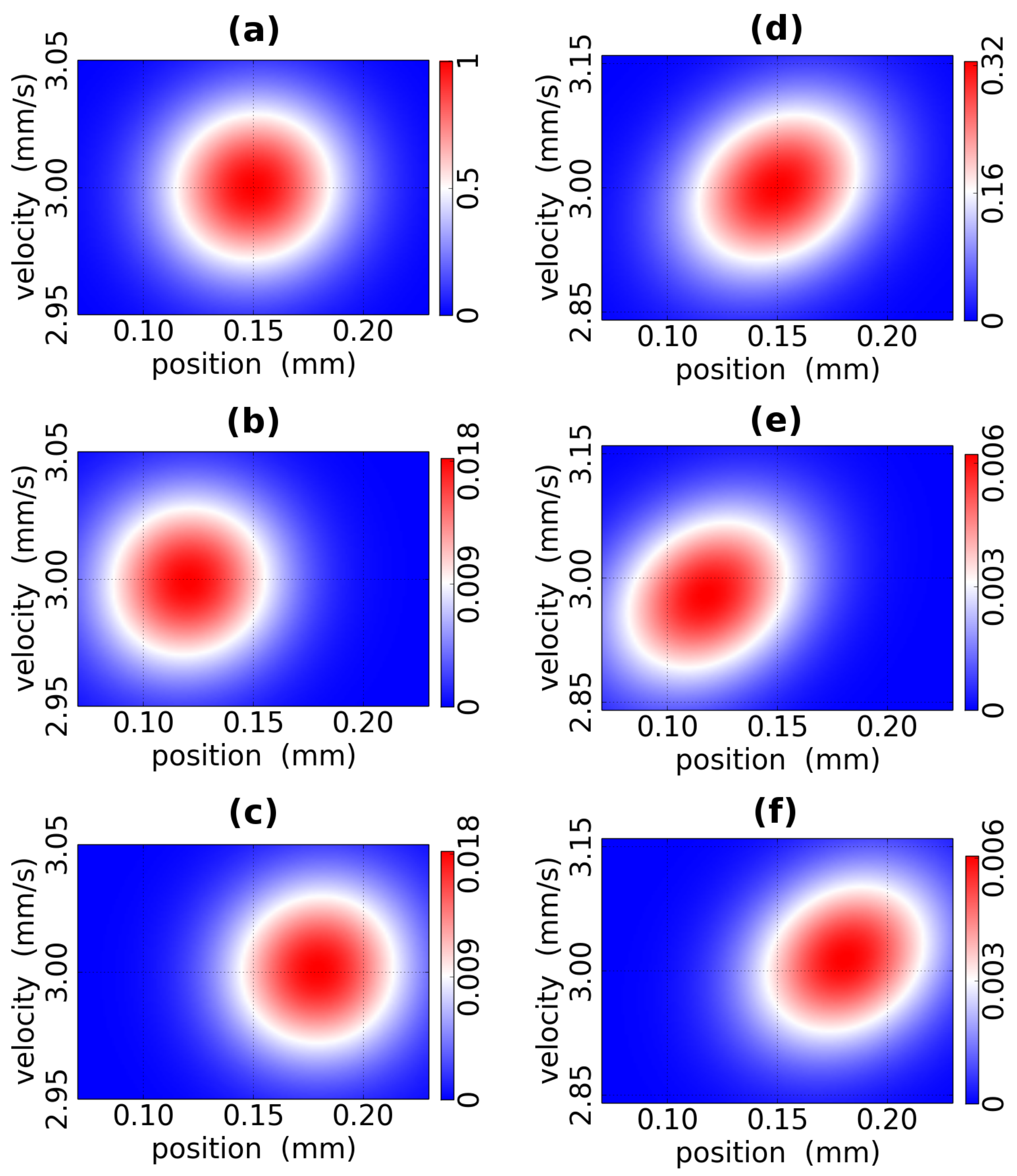}
\caption{Husimi distributions for a $^{87}$Rb atom in a pure (a-c) and
  a mixed, finite-temperature (d-f) state, i.e., $H_{\alpha_0, x_0,
    v_0}^{(t)}(\tilde{x}, \tilde{v})$ and $\mathcal{H}_{\alpha_0, x_0,
    v_0, \Delta v}^{(t)}(\tilde{x}, \tilde{v})$ respectively. System
  parameters are: $m = 86.909$~u, $\sigma = 30$~$\mu$m, $x_0 =
  -0.15$~mm, $v_0 = 3$~mm/s, $\Delta v = 0.1$~mm/s, and $t =
  100$~ms. The aperture function is defined by
  Eq.~(\ref{chi-shift}). $\gamma = 0$ in (a) and (d), $\gamma =
  100$~s$^{-1}$ in (b) and (e), and $\gamma = -100$~s$^{-1}$ in (c)
  and (f).}
\label{fig3}
\end{figure}

In the semiclassical regime, the shape of the transmitted WP depends
predominantly on the form of the aperture function $\chi_{\tau}$ in
the vicinity of the time $t_0$, at which the classical particle
crosses the barrier, and is largely insensitive to the behavior of
$\chi_{\tau}$ close to the ends of the time interval $0 \leq \tau \leq
t$. So, in order to increase the overall transmission probability we
consider the aperture function (see Fig.~\ref{fig2}(a))
\begin{equation}
  \chi_{\tau} = \min \left\{ e^{\gamma (\tau - t_1)} \,, \; 1 \right\}
  \,, \quad t_1 = \left\{
  \begin{array}{cl}
    \frac{3 t_0}{2} & \mathrm{if} \; \gamma > 0 \\[0.2cm]
    \frac{t_0}{2} & \mathrm{if} \; \gamma \leq 0
  \end{array} \right.
\label{chi-shift}
\end{equation}
when numerically evaluating the integrals in Eqs.~(\ref{husimi}) and
(\ref{HUSIMI_temperature_2}), instead of the one given by
Eq.~(\ref{chi_exp}). Figure~\ref{fig3} shows the corresponding Husimi
distributions $H_{\alpha_0, x_0, v_0}^{(t)}(\tilde{x},\tilde{v})$ and
$\mathcal{H}_{\alpha_0, x_0, v_0, \Delta
  v}^{(t)}(\tilde{x},\tilde{v})$ as functions of position $\tilde{x}$
and velocity $\tilde{v}$ for different values of
$\gamma$. Figure~\ref{fig3}(a-c) represent the pure state case, and
Fig.~\ref{fig3}(d-f) correspond to the case of a mixed,
finite-temperature state. The spatial shift of the Husimi distribution
is well pronounced in the figure, and its numerical value is found to
be in good agreement with the predictions of Eq.~(\ref{shift}), i.e.,
$\Delta x = \mp 30$~$\mu$m for $\gamma = \pm 100$~s$^{-1}$. It is
interesting to observe a slight change of the average velocity of the
particle in the mixed state case (see Fig.~\ref{fig3}(e,f)). This
velocity shift stems from the fact that WPs with different average
velocities, comprising the mixed state, arrive at the barrier at
different times and, as a result, are subject to different values of
the transparency function. As we show later, this effect can be
exploited to reduce the phase-space uncertainty of (and effectively
cool down) an atomic cloud.

\begin{figure}[h]
\includegraphics[width=3.2in]{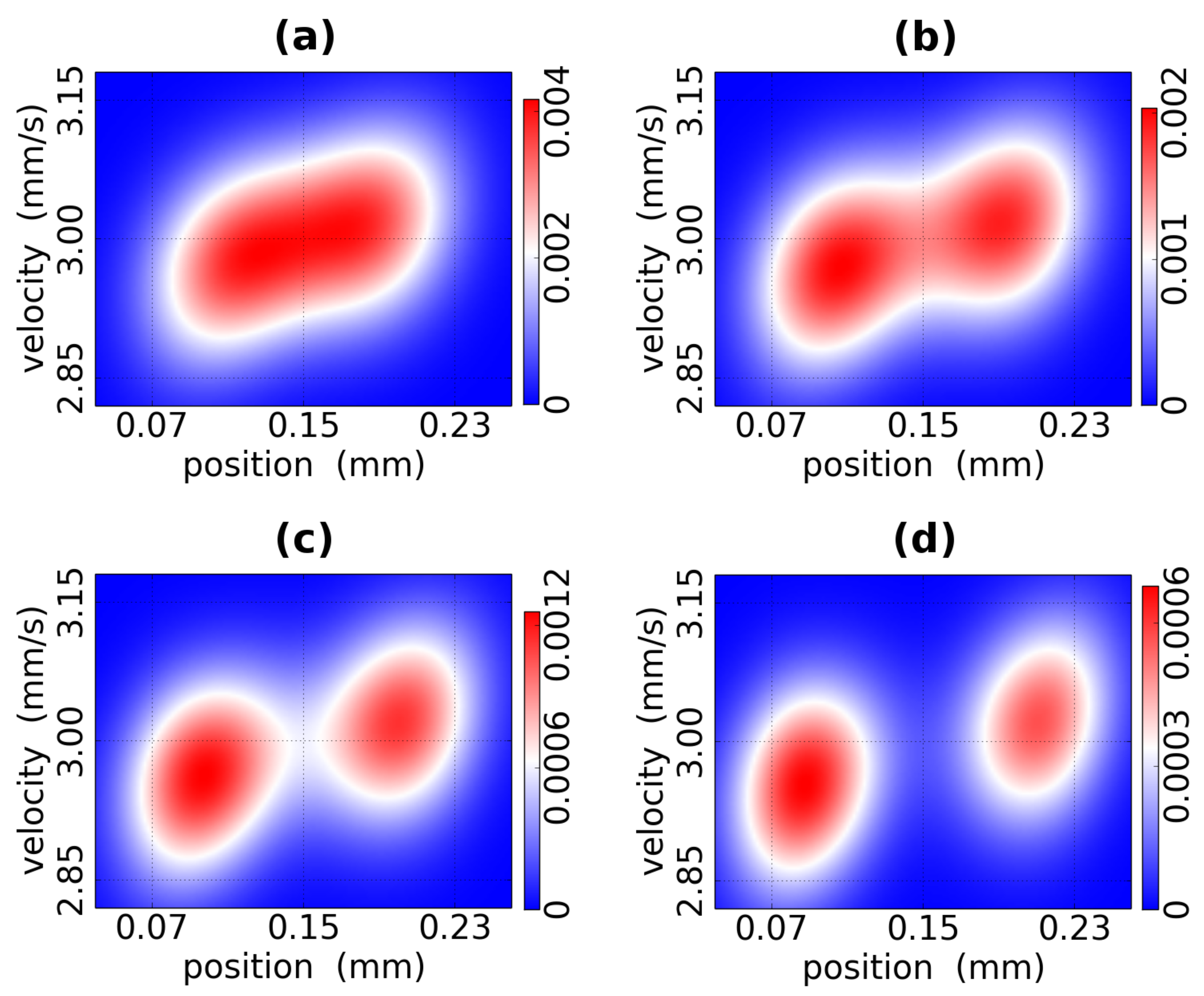}
\caption{Husimi distribution $\mathcal{H}_{\alpha_0, x_0, v_0, \Delta
    v}^{(t)}(\tilde{x}, \tilde{v})$ for a mixed state, characterized
  by the same set of parameters as in Fig.~\ref{fig3}(d-f), in the
  presence of an absorbing barrier specified by Eq.~(\ref{chi-split})
  with (a) $\gamma = 125$~s$^{-1}$, (b) $\gamma = 150$~s$^{-1}$, (c)
  $\gamma = 175$~s$^{-1}$, and (d) $\gamma = 225$~s$^{-1}$.}
\label{fig4}
\end{figure}

\subsection{Splitting}
\label{sec:split}

We now consider a different scenario in which the aperture function
$\chi_{\tau}$ in the vicinity of $t_0$ is given by an equally weighted
sum of an increasing and a decaying exponential, $e^{\gamma \tau}$ and
$e^{-\gamma \tau}$. As before, in order to increase the overall
transmission probability, we take $\chi_{\tau}$ to be unity around the
ends of the interval $0 \leq \tau \leq t$. Thus, we choose (see
Fig.~\ref{fig2}(b))
\begin{equation}
  \chi_{\tau} = \min \left\{ \frac{\cosh[\gamma (\tau -
      t_0)]}{\cosh(\gamma t_0 / 2)} \,, \; 1 \right\} \,.
\label{chi-split}
\end{equation}
Figure~\ref{fig4} shows the response of a finite-temperature WP,
characterized by the same set of parameters as above, to an absorbing
barrier specified by Eq.~(\ref{chi-split}). As the rate $\gamma$
increases, the WP stretches and eventually splits in two practically
non-overlapping parts. A slight difference between the average
velocities of the two parts has the same physical origin as in the
shifting scenario (see Fig.~\ref{fig3}(e,f)).

The absorption-based WP splitting mechanism presented here may be
utilized in designing new types of matter-wave
interferometers. Indeed, the two WPs produced by the splitting
continue propagating along the same path in the coordinate space. If
this path traverses a region with an external potential that is
nonuniform in space and time, such as a time-dependent disorder, then
the two WPs will accumulate different phases in the course of their
motion, and their subsequent recombination will give rise to an
interference pattern. The interference pattern can subsequently be
used to extract information about the potential.

\begin{figure}[h]
\includegraphics[width=3.4in]{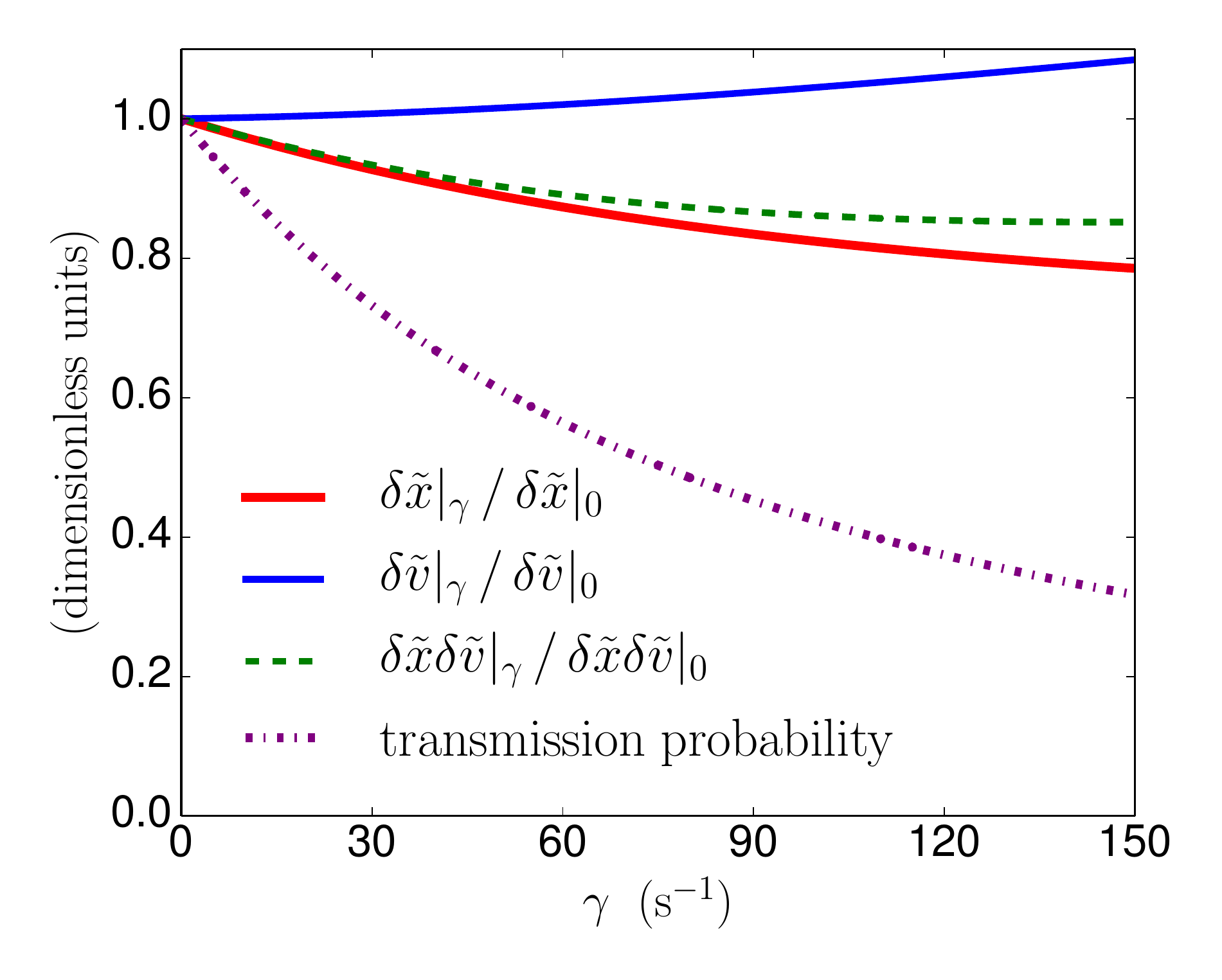}
\caption{Relative spatial (solid red curve), velocity (solid blue
  curve) and phase-space (dashed green curve) uncertainties, and total
  transmission probability (dashed-dotted purple curve) as functions
  of the rate $\gamma$ for the aperture function defined in
  Eq.~(\ref{chi-squeeze}). All system parameters are the same as in
  Fig.~\ref{fig3}(d-f) and Fig.~\ref{fig4}.}
\label{fig5}
\end{figure}

\subsection{Squeezing and cooling}
\label{sec:squeeze}

Finally, we consider a scenario in which the barrier first opens
exponentially until the time $t_0$ and then closes exponentially, so
that the aperture function reads (see Fig.~\ref{fig2}(c))
\begin{equation}
  \chi_{\tau} = e^{-\gamma |\tau - t_0|} \,, \quad \gamma > 0 \,.
\label{chi-squeeze}
\end{equation}
In this case, the Husimi distribution $\mathcal{H}_{\alpha_0, x_0,
  v_0, \Delta v}^{(t)}(\tilde{x}, \tilde{v})$ of a transmitted WP for
$\gamma > 0$ appears to be squeezed in the $\tilde{x}$ direction and
stretched in the $\tilde{v}$ direction compared to the free-particle
case. Figure~\ref{fig5} shows the $\gamma$-dependence of position
(solid red curve) and velocity (solid blue curve) dispersions of the
WP, $\delta\tilde{x}$ and $\delta\tilde{v}$ respectively, computed
with respect to the Husimi distribution:
\begin{equation}
  \delta (\cdot) = \left[ \int d\tilde{x} d\tilde{v} (\cdot)^2
    \mathcal{H} - \left( \int d\tilde{x} d\tilde{v} (\cdot)
    \mathcal{H} \right)^2 \right]^{1/2} \,.
\label{disp_hus}
\end{equation}
The initial WP is characterized by the same set of parameters as
above.

It is interesting to observe that as $\gamma$ grows the decrease of
$\delta \tilde{x}$ occurs at a higher rate than the increase of
$\delta \tilde{v}$. For instance, at $\gamma = 150$~s$^{-1}$ the
spatial dispersion is reduced by over 20\% compared to its value in
the absence of a barrier, whereas the corresponding relative increase
in the velocity dispersion is less than 10\%. This means that the
overall phase-space uncertainty $\delta \tilde{x} (m \delta
\tilde{v})$ decreases with growing $\gamma$. The $\gamma$-dependence
of the relative phase-space uncertainty is shown by a dashed green
curve in Fig.~\ref{fig5}. For the given set of parameters, $\delta
\tilde{x} (m \delta \tilde{v}) \big|_{\gamma = 0} \simeq 3.12 \hbar$,
whereas $\delta \tilde{x} (m \delta \tilde{v}) \big|_{\gamma = 150 \,
  \mathrm{s}^{-1}} \simeq 2.66 \hbar$. (We note that the Heisenberg
uncertainty principle, with all averages computed with respect to a
Husimi distribution function, states that $\delta \tilde{x} (m \delta
\tilde{v}) \geq \hbar$ \cite{Bal14Quantum}.) In other words, the
velocity spread of the WP at a finite $\gamma$ is closer to the
Heisenberg limit $\delta \tilde{v}_H = \hbar / \delta \tilde{x}$ than
that of the corresponding free-particle WP. This in turn means that
the moving particle gets effectively cooled down by the absorbing
barrier. The cooling occurs through absorption of those components of
the mixed state that have the largest deviations of the velocity from
its average value. (The dotted-dashed purple curve in Fig.~\ref{fig5}
shows the decay of the overall transmission probability defined as
$\int_0^{+\infty} d\tilde{x} \int_{-\infty}^{+\infty} d\tilde{v} \,
\mathcal{H}$.) In a sense, the effect is similar in nature to that of
evaporative cooling \cite{AEM+95Observation}.


\section{Conclusion}
\label{sec:conclusion}

In summary, we have demonstrated that a moving WP of quantum matter
can be flexibly manipulated with the help of a thin stationary
absorbing barrier whose transparency changes in time according to an
externally prescribed protocol. In particular, the WP transmitted
through the barrier may be spatially shifted, split in two, or
squeezed and cooled compared to the corresponding WP in free
space. The reported effects can be observed in a laboratory setting
using a cloud of ultra-cold atoms akin to that produced in experiments
in Ref.~\cite{JMR+12Coherent} and a laser light sheet of variable
intensity.

In this paper, being mainly interested in a proof-of-principle
demonstration of absorption-based WP control, we have only considered
barrier apertures of relatively simple, compact functional forms. In
real world situations however aperture function optimization could be
used to steer the wave function into a desired target state. (See
Ref.~\cite{NJF+10Dynamic} for an example of the optimization approach
in the context of control of atomic WPs in atom chips.) Other
important extensions of the present work would be to generalize our
theory to the case of interacting particles and to investigate if
there are any new effects produced by a time-dependent absorbing
barrier of a finite spatial extent.

We believe that our findings may become of considerable value in areas
of physics concerned with matter-wave interferometry, quantum control,
and quantum metrology, as well as facilitate better understanding of
effects of absorption in quantum systems.

\begin{acknowledgments}
  The author thanks Ilya Arakelyan, Maximilien Barbier, and Adolfo del
  Campo for valuable comments and stimulating discussions, and
  acknowledges the financial support of EPSRC under Grant
  No.~EP/K024116/1.
\end{acknowledgments}


\appendix

\section{Evaluation of $\Phi_{\alpha_0, \tilde{x}, \tilde{v}, x_0, v_0, \Delta
    v}^{(t,\tau,\tau')}$}
\label{app1}

Here we derive a closed form expression for the function
$\Phi_{\alpha_0, \tilde{x}, \tilde{v}, x_0, v_0, \Delta
  v}^{(t,\tau,\tau')}$ defined by Eq.~(\ref{Phi_function}).

Using Eq.~(\ref{psi_free_flight}), we write
\begin{widetext}
\begin{align}
  \psi_{\alpha_0, x_0, v}^{(\tau)}(0) &= \left( \frac{2
    \alpha_{\tau}^2}{\pi \alpha_0} \right)^{1/4} \exp \left(
  -\alpha_{\tau} (x_0 + v \tau)^2 - i \frac{m v}{\hbar} (x_0 + v \tau)
  + i \frac{m v^2 \tau}{2 \hbar} \right) \nonumber \\ &= \left(
  \frac{2 \alpha_{\tau}^2}{\pi \alpha_0} \right)^{1/4} \exp \left[
    -\left( \alpha_{\tau} \tau + i \frac{m}{2 \hbar} \right) \tau v^2
    - 2 \left( \alpha_{\tau} \tau + i \frac{m}{2 \hbar} \right) x_0 v
    - \alpha_{\tau} x_0^2 \right] \nonumber \\ &= \left( \frac{2
    \alpha_{\tau}^2}{\pi \alpha_0} \right)^{1/4} \exp \left( -i
  \frac{m \tau \alpha_{\tau}}{2 \hbar \alpha_0} v^2 - i \frac{m x_0
    \alpha_{\tau}}{\hbar \alpha_0} v - \alpha_{\tau} x_0^2 \right)
  \nonumber \,.
\end{align}
\end{widetext}
In the last line we have used the identity $\alpha_{\tau} \tau +
\frac{i m}{2 \hbar} = i \frac{m \alpha_{\tau}}{2 \hbar
  \alpha_0}$. Similarly, we have
\begin{align}
  \left[ \psi_{\alpha_0, x_0, v}^{(\tau')}(0) \right]^* = &\left(
  \frac{2 (\alpha_{\tau'}^*)^2}{\pi \alpha_0} \right)^{1/4}
  \nonumber\\ &\times \exp \left( i \frac{m \tau' \alpha_{\tau'}^*}{2
    \hbar \alpha_0} v^2 + i \frac{m x_0 \alpha_{\tau'}^*}{\hbar
    \alpha_0} v - \alpha_{\tau'}^* x_0^2 \right) \,. \nonumber
\end{align}
Therefore
\begin{align}
  e^{-(v - v_0)^2 / \Delta v^2} &\psi_{\alpha_0, x_0, v}^{(\tau)}(0)
  \left[ \psi_{\alpha_0, x_0, v}^{(\tau')}(0) \right]^* \nonumber\\ &=
  \sqrt{\frac{2 R \alpha_0}{\pi}} \exp \left( -A v^2 + B v - C \right)
  \,, \nonumber
\end{align}
where
\begin{align}
  A &= \frac{1}{\Delta v^2} + i \frac{m \left( \tau \alpha_{\tau} -
    \tau' \alpha_{\tau'}^* \right)}{2 \hbar \alpha_0} \,, \nonumber
  \\ B &= \frac{2 v_0}{\Delta v^2} - i \frac{m x_0 \left(
    \alpha_{\tau} - \alpha_{\tau'}^* \right)}{\hbar \alpha_0} \,,
  \nonumber \\ C &= \frac{v_0^2}{\Delta v^2} + \left( \alpha_{\tau} +
  \alpha_{\tau'}^* \right) x_0^2 \,, \nonumber \\ R &=
  \frac{\alpha_{\tau} \alpha_{\tau'}^*}{\alpha_0^2} \,. \nonumber
\end{align}
Also, using $\frac{v}{\alpha_{|x_0|/v}} = \frac{v}{\alpha_0} + i
\frac{2 \hbar |x_0|}{m} = \frac{v}{\alpha_0} - i \frac{2 \hbar
  x_0}{m}$, we write
\begin{widetext}
\begin{align}
  \left( \frac{\alpha_{t-\tau} \tilde{v}}{\alpha_{\tilde{t}}} +
  \frac{\alpha_{\tau} v}{\alpha_{|x_0|/v}} \right) \left(
  \frac{\alpha_{t-\tau'} \tilde{v}}{\alpha_{\tilde{t}}} +
  \frac{\alpha_{\tau'} v}{\alpha_{|x_0|/v}} \right)^{\!*} &= \left(
  \frac{\alpha_{t-\tau} \tilde{v}}{\alpha_{\tilde{t}}} - i \frac{2
    \hbar x_0 \alpha_{\tau}}{m} + \frac{\alpha_{\tau}}{\alpha_0} v
  \right) \left( \frac{\alpha_{t-\tau'}^*
    \tilde{v}}{\alpha_{\tilde{t}}^*} + i \frac{2 \hbar x_0
    \alpha_{\tau'}^*}{m} + \frac{\alpha_{\tau'}^*}{\alpha_0} v \right)
  \nonumber\\ &= R v^2 + S v + T \,, \nonumber
\end{align}
\end{widetext}
where
\begin{align}
  S &= \frac{\tilde{v}}{\alpha_0} \left( \frac{\alpha_{\tau}
    \alpha_{t-\tau'}^*}{\alpha_{\tilde{t}}^*} + \frac{\alpha_{t-\tau}
    \alpha_{\tau'}^*}{\alpha_{\tilde{t}}} \right) \,, \nonumber \\ T
  &= \left( \frac{\alpha_{t-\tau} \tilde{v}}{\alpha_{\tilde{t}}} - i
  \frac{2 \hbar x_0 \alpha_{\tau}}{m} \right) \left(
  \frac{\alpha_{t-\tau'}^* \tilde{v}}{\alpha_{\tilde{t}}^*} + i
  \frac{2 \hbar x_0 \alpha_{\tau'}^*}{m} \right) \,. \nonumber
\end{align}
Substituting the above expressions into Eq.~(\ref{Phi_function}), we
obtain
\begin{widetext}
\begin{align}
  \Phi_{\alpha_0, \tilde{x}, \tilde{v}, x_0, v_0, \Delta
    v}^{(t,\tau,\tau')} &= \frac{\sqrt{2 R \alpha_0}}{\pi \Delta v}
  \int\limits_{-\infty}^{+\infty} \, dv \left( R v^2 + S v + T \right)
  \exp \left( -A v^2 + B v - C \right) \nonumber \\ &= \frac{1}{\Delta
    v} \sqrt{\frac{2 R \alpha_0}{\pi}} \left( \frac{T}{A^{1/2}} +
  \frac{B S + R}{2 A^{3/2}} + \frac{B^2 R}{4 A^{5/2}} \right) \exp
  \left( \frac{B^2}{4 A} - C \right) \,. \nonumber
\end{align}
\end{widetext}
The last expression can be directly adopted for numerical evaluation
of the thermal-state Husimi distribution.

Finally, we note that, as expected, the last expression respects the
identity
\begin{align}
  \lim_{\Delta v \rightarrow 0} &\Phi_{\alpha_0, \tilde{x}, \tilde{v},
    x_0, v_0, \Delta v}^{(t,\tau,\tau')} \nonumber\\ &\quad = \left(
  \frac{\alpha_{t-\tau} \tilde{v}}{\alpha_{\tilde{t}}} +
  \frac{\alpha_{\tau} v_0}{\alpha_{t_0}} \right) \left(
  \frac{\alpha_{t-\tau'} \tilde{v}}{\alpha_{\tilde{t}}} +
  \frac{\alpha_{\tau'} v_0}{\alpha_{t_0}} \right)^{\!*}
  \nonumber\\ &\quad\quad \times \psi_{\alpha_0, x_0, v_0}^{(\tau)}(0)
  \left[ \psi_{\alpha_0, x_0, v_0}^{(\tau')}(0) \right]^* \nonumber
\end{align}
with $t_0 = |x_0| / v_0$, thus recovering
\begin{equation}
  \lim_{\Delta v \rightarrow 0} \mathcal{H}_{\alpha_0, x_0, v_0,
    \Delta v}^{(t)}(\tilde{x}, \tilde{v}) = H_{\alpha_0, x_0,
    v_0}^{(t)}(\tilde{x}, \tilde{v}) \,. \nonumber
\end{equation}

\section{Peak of $ H_{\alpha_0, x_0,
    v_0}^{(t)}(\tilde{x}, \tilde{v})$ for $\chi_{\tau} = \chi_0
  e^{\gamma \tau}$ in the semiclassical regime}
\label{app2}

Here we provide a derivation of Eq.~(\ref{shift}).

In the semiclassical regime, $1 \ll \frac{|x_0|}{\sigma} \lesssim
\frac{v_0 t}{2 \sigma} \ll \frac{\sigma}{2 \lambdabar}$ with
$\lambdabar = \hbar / (m v_0)$, we define a small parameter
\begin{equation}
  \epsilon = \frac{\hbar t}{m \sigma^2} \ll 1 \,. \nonumber
\end{equation}
($\epsilon$ plays the role of an effective Planck's constant.) Using
$\alpha_{\tau} = \alpha_0 \left(1 + i \frac{\tau}{t} \epsilon
\right)^{-1} = \alpha_0 + \mathcal{O}(\epsilon)$ for $0 < \tau < t$,
we write
\begin{widetext}
\begin{align}
  \psi_{\alpha_0, x_0, v_0}^{(\tau)}(0) &= \left( \frac{2
    \alpha_{\tau}^2}{\pi \alpha_0} \right)^{1/4} \exp \left(
  -\alpha_{\tau} (x_0 + v_0 \tau)^2 - i \frac{m v_0}{\hbar} (x_0 + v_0
  \tau) + i \frac{m v_0^2 \tau}{2 \hbar} \right) \nonumber \\ &=
  \left[ \left( \frac{2 \alpha_0}{\pi} \right)^{1/4} +
    \mathcal{O}(\epsilon) \right] \exp \left( -\alpha_0 (x_0 + v_0
  \tau)^2 - i \frac{m v_0}{\hbar} (x_0 + v_0 \tau) + i \frac{m v_0^2
    \tau}{2 \hbar} + \mathcal{O}(\epsilon) \right) \nonumber
  \\ &\simeq \left( \frac{2 \alpha_0}{\pi} \right)^{1/4} \exp \left[
    -\alpha_0 v_0^2 \tau^2 - \left( 2 \alpha_0 x_0 v_0 + i \frac{m
      v_0^2}{2 \hbar} \right) \tau - \alpha_0 x_0^2 - i \frac{m x_0
      v_0}{\hbar} \right] \,. \nonumber
\end{align}
Similarly,
\begin{align}
  &\left[ \psi_{\alpha_0, x_0, v_0}^{(\tau')}(0) \right]^* \simeq
  \left( \frac{2 \alpha_0}{\pi} \right)^{1/4} \exp \left[ -\alpha_0
    v_0^2 \tau'^2 - \left( 2 \alpha_0 x_0 v_0 - i \frac{m v_0^2}{2
      \hbar} \right) \tau' - \alpha_0 x_0^2 + i \frac{m x_0
      v_0}{\hbar} \right] \,, \nonumber \\ &\left[ \psi_{\alpha_0,
      \tilde{x}, \tilde{v}}^{(\tau-t)}(0) \right]^* \simeq \left(
  \frac{2 \alpha_0}{\pi} \right)^{1/4} \exp \left[ -\alpha_0
    \tilde{v}^2 (\tau - t)^2 - \left( 2 \alpha_0 \tilde{x} \tilde{v} -
    i \frac{m \tilde{v}^2}{2 \hbar} \right) (\tau - t) - \alpha_0
    \tilde{x}^2 + i \frac{m \tilde{x} \tilde{v}}{\hbar} \right] \,,
  \nonumber \\ &\psi_{\alpha_0, \tilde{x}, \tilde{v}}^{(\tau'-t)}(0)
  \simeq \left( \frac{2 \alpha_0}{\pi} \right)^{1/4} \exp \left[
    -\alpha_0 \tilde{v}^2 (\tau' - t)^2 - \left( 2 \alpha_0 \tilde{x}
    \tilde{v} + i \frac{m \tilde{v}^2}{2 \hbar} \right) (\tau' - t) -
    \alpha_0 \tilde{x}^2 - i \frac{m \tilde{x} \tilde{v}}{\hbar}
    \right] \,.  \nonumber
\end{align}
\end{widetext}
Then,
\begin{align}
  &\left[ \psi_{\alpha_0, \tilde{x}, \tilde{v}}^{(\tau-t)}(0)
    \right]^* \psi_{\alpha_0, \tilde{x}, \tilde{v}}^{(\tau'-t)}(0)
  \,\psi_{\alpha_0, x_0, v_0}^{(\tau)}(0) \left[ \psi_{\alpha_0, x_0,
      v_0}^{(\tau')}(0) \right]^* \nonumber\\ &\qquad \simeq \frac{2
    \alpha_0}{\pi} \exp \big[ -U \left( \tau^2 + \tau'^2 \right) -
    \left( V_R - i V_I \right) \tau
    \nonumber\\ &\qquad\qquad\qquad\qquad -\left( V_R + i V_I \right)
    \tau' - W \big] \,, \nonumber
\end{align}
where
\begin{align}
  &U = \alpha_0 \left( \tilde{v}^2 + v_0^2 \right) \,, \nonumber
  \\ &V_R = 2 \alpha_0 \big[ (\tilde{x} - \tilde{v} t) \tilde{v} + x_0
    v_0 \big] \,, \nonumber \\ &V_I = \frac{m \left( \tilde{v}^2 -
    v_0^2 \right)}{2 \hbar} \,, \nonumber \\ &W = 2 \alpha_0 \left[
    \left( \tilde{x} - \tilde{v} t \right)^2 + x_0^2 \right]
  \,. \nonumber
\end{align}

The Husimi distribution now reads
\begin{align}
  &H_{\alpha_0, x_0, v_0}^{(t)}(\tilde{x}, \tilde{v}) \simeq
  \frac{(\tilde{v} + v_0)^2}{4} \int\limits_0^t d\tau \int\limits_0^t
  d\tau' \, \chi_{\tau} \chi_{\tau'} \nonumber\\ &\;\; \times \left[
    \psi_{\alpha_0, \tilde{x}, \tilde{v}}^{(\tau-t)}(0) \right]^*
  \psi_{\alpha_0, \tilde{x}, \tilde{v}}^{(\tau'-t)}(0)
  \,\psi_{\alpha_0, x_0, v_0}^{(\tau)}(0) \left[ \psi_{\alpha_0, x_0,
      v_0}^{(\tau')}(0) \right]^* \nonumber \\ &= \frac{\alpha_0
    (\tilde{v} + v_0)^2}{2 \pi} e^{-W} \left| \int\limits_0^t d\tau \,
  \chi_{\tau} e^{-U \tau^2 - (V_R + i V_I) \tau} \right|^2 . \nonumber
\end{align}
Taking $\chi_\tau = \chi_0 e^{\gamma \tau}$, we get
\begin{align}
  H_{\alpha_0, x_0, v_0}^{(t)}&(\tilde{x}, \tilde{v}) \simeq \chi_0^2
  \, \frac{\alpha_0 (\tilde{v} + v_0)^2}{2 \pi} e^{-W}
  \nonumber\\ &\times \left| \int\limits_0^t d\tau \, e^{-U \tau^2 -
    (V_R - \gamma + i V_I) \tau} \right|^2 . \nonumber
\end{align}
The evaluation of the last integral substantially simplifies if we
consider the position $\tilde{x}$ to lie sufficiently close to the
point $x_0 + v_0 t$ and the velocity $\tilde{v}$ to be close to
$v_0$. In this case, the main contribution to the integral comes from
the time interval $t_0 + \delta t - \sigma / v_0 \lesssim \tau
\lesssim t_0 + \delta t + \sigma / v_0$ with $t_0 = \frac{|x_0|}{v_0}$
and $\delta t = \frac{\gamma \sigma^2}{2 v_0^2}$. (Indeed, since $U
\simeq 2 \alpha_0 v_0^2$, $V_R \simeq 4 \alpha_0 x_0 v_0$ and $V_I
\simeq 0$, the exponent $-U \tau^2 - (V_R - \gamma) \tau$ peaks at
$\tau_{\max} = \frac{-V_R + \gamma}{2 U} \simeq \frac{|x_0|}{v_0} +
\frac{\gamma}{4 \alpha_0 v_0^2} = t_0 + \delta t$. The width of the
peak can be estimated as $U^{-1/2} \simeq \frac{1}{\sqrt{2 \alpha_0
    v_0^2}} = \frac{\sigma}{v_0}.$) This interval is contained well
inside the integration range $0 < \tau < t$, provided that $|\delta t|
\ll t_0$ or, equivalently,
\begin{equation}
  |\gamma| \ll \frac{2 |x_0| v_0}{\sigma^2} \,. \nonumber
\end{equation}
Then,
\begin{align}
  &H_{\alpha_0, x_0, v_0}^{(t)}(\tilde{x}, \tilde{v}) \simeq \chi_0^2
  \, \frac{\alpha_0 (\tilde{v} + v_0)^2}{2 \pi} e^{-W}
  \nonumber\\ &\qquad\qquad \times \left| \,
  \int\limits_{-\infty}^{+\infty} d\tau \, e^{-U \tau^2 - (V_R -
    \gamma + i V_I) \tau} \right|^2 \nonumber \\ &= \chi_0^2
  \frac{\alpha_0 (\tilde{v} + v_0)^2}{2 U} \exp \left( \frac{(V_R -
    \gamma)^2 - V_I^2}{2 U} - W \right) . \nonumber
\end{align}

As we are only interested in the form of the Husimi distribution in
the vicinity of the phase-space point $(\tilde{x}, \tilde{v}) = (x_0 +
v_0 t, v_0)$, the exponential prefactor can be approximated by
$\chi_0^2$, yielding
\begin{equation}
  H_{\alpha_0, x_0, v_0}^{(t)}(\tilde{x}, \tilde{v}) \simeq \chi_0^2
  \, e^{\Xi} \nonumber
\end{equation}
with
\begin{equation}
  \Xi = \frac{(V_R - \gamma)^2 - V_I^2}{2 U} - W \,. \nonumber
\end{equation}
It is now straightforward (although tedious) to show that the exponent
$\Xi$ (and so the Husimi distribution) has a local maximum at the
phase-space point $(\tilde{x}_{\max}, \tilde{v}_{\max}) = (x_0 + v_0 t
+ \Delta x, v_0)$, where
\begin{equation}
  \Delta x = -\frac{\gamma}{2 \alpha_0 v_0} = -\frac{\gamma
    \sigma^2}{v_0} \,. \nonumber
\end{equation}
Indeed, one can verify that
\begin{equation}
  \left. \frac{\partial \Xi}{\partial \tilde{x}}
  \right|_{(\tilde{x}_{\max}, \tilde{v}_{\max})} =
  \left. \frac{\partial \Xi}{\partial \tilde{v}}
  \right|_{(\tilde{x}_{\max}, \tilde{v}_{\max})} = 0 \,, \nonumber
\end{equation}
\begin{equation}
  \left. \det \left(
  \begin{array}{cc}
    \frac{\partial^2 \Xi}{\partial \tilde{x}^2} & \frac{\partial^2
      \Xi}{\partial \tilde{x} \partial \tilde{v}} \\[0.2cm]
    \frac{\partial^2 \Xi}{\partial \tilde{v} \partial \tilde{x}} &
    \frac{\partial^2 \Xi}{\partial \tilde{v}^2}
  \end{array} \right) \right|_{(\tilde{x}_{\max}, \tilde{v}_{\max})}
  = \frac{m^2}{\hbar^2} > 0 \,, \nonumber
\end{equation}
and
\begin{equation}
  \left. \frac{\partial^2 \Xi}{\partial \tilde{x}^2}
  \right|_{(\tilde{x}_{\max}, \tilde{v}_{\max})} = -2 \alpha_0 < 0 \,.
  \nonumber
\end{equation}

%


\begin{thebibliography}{35}%
\makeatletter
\providecommand \@ifxundefined [1]{%
 \@ifx{#1\undefined}
}%
\providecommand \@ifnum [1]{%
 \ifnum #1\expandafter \@firstoftwo
 \else \expandafter \@secondoftwo
 \fi
}%
\providecommand \@ifx [1]{%
 \ifx #1\expandafter \@firstoftwo
 \else \expandafter \@secondoftwo
 \fi
}%
\providecommand \natexlab [1]{#1}%
\providecommand \enquote  [1]{``#1''}%
\providecommand \bibnamefont  [1]{#1}%
\providecommand \bibfnamefont [1]{#1}%
\providecommand \citenamefont [1]{#1}%
\providecommand \href@noop [0]{\@secondoftwo}%
\providecommand \href [0]{\begingroup \@sanitize@url \@href}%
\providecommand \@href[1]{\@@startlink{#1}\@@href}%
\providecommand \@@href[1]{\endgroup#1\@@endlink}%
\providecommand \@sanitize@url [0]{\catcode `\\12\catcode `\$12\catcode
  `\&12\catcode `\#12\catcode `\^12\catcode `\_12\catcode `\%12\relax}%
\providecommand \@@startlink[1]{}%
\providecommand \@@endlink[0]{}%
\providecommand \url  [0]{\begingroup\@sanitize@url \@url }%
\providecommand \@url [1]{\endgroup\@href {#1}{\urlprefix }}%
\providecommand \urlprefix  [0]{URL }%
\providecommand \Eprint [0]{\href }%
\providecommand \doibase [0]{http://dx.doi.org/}%
\providecommand \selectlanguage [0]{\@gobble}%
\providecommand \bibinfo  [0]{\@secondoftwo}%
\providecommand \bibfield  [0]{\@secondoftwo}%
\providecommand \translation [1]{[#1]}%
\providecommand \BibitemOpen [0]{}%
\providecommand \bibitemStop [0]{}%
\providecommand \bibitemNoStop [0]{.\EOS\space}%
\providecommand \EOS [0]{\spacefactor3000\relax}%
\providecommand \BibitemShut  [1]{\csname bibitem#1\endcsname}%
\let\auto@bib@innerbib\@empty
\bibitem [{\citenamefont {Cronin}\ \emph {et~al.}(2009)\citenamefont {Cronin},
  \citenamefont {Schmiedmayer},\ and\ \citenamefont {Pritchard}}]{CSP09Optics}%
  \BibitemOpen
  \bibfield  {author} {\bibinfo {author} {\bibfnamefont {A.~D.}\ \bibnamefont
  {Cronin}}, \bibinfo {author} {\bibfnamefont {J.}~\bibnamefont
  {Schmiedmayer}}, \ and\ \bibinfo {author} {\bibfnamefont {D.~E.}\
  \bibnamefont {Pritchard}},\ }\href@noop {} {\bibfield  {journal} {\bibinfo
  {journal} {Rev. Mod. Phys.}\ }\textbf {\bibinfo {volume} {81}},\ \bibinfo
  {pages} {1051} (\bibinfo {year} {2009})}\BibitemShut {NoStop}%
\bibitem [{\citenamefont {Hornberger}\ \emph {et~al.}(2012)\citenamefont
  {Hornberger}, \citenamefont {Gerlich}, \citenamefont {Haslinger},
  \citenamefont {Nimmrichter},\ and\ \citenamefont {Arndt}}]{HGH+12Colloquium}%
  \BibitemOpen
  \bibfield  {author} {\bibinfo {author} {\bibfnamefont {K.}~\bibnamefont
  {Hornberger}}, \bibinfo {author} {\bibfnamefont {S.}~\bibnamefont {Gerlich}},
  \bibinfo {author} {\bibfnamefont {P.}~\bibnamefont {Haslinger}}, \bibinfo
  {author} {\bibfnamefont {S.}~\bibnamefont {Nimmrichter}}, \ and\ \bibinfo
  {author} {\bibfnamefont {M.}~\bibnamefont {Arndt}},\ }\href@noop {}
  {\bibfield  {journal} {\bibinfo  {journal} {Rev. Mod. Phys.}\ }\textbf
  {\bibinfo {volume} {84}},\ \bibinfo {pages} {157} (\bibinfo {year}
  {2012})}\BibitemShut {NoStop}%
\bibitem [{\citenamefont {Arndt}(2014)}]{Arn14DeBroglie}%
  \BibitemOpen
  \bibfield  {author} {\bibinfo {author} {\bibfnamefont {M.}~\bibnamefont
  {Arndt}},\ }\href@noop {} {\bibfield  {journal} {\bibinfo  {journal} {Phys.
  Today}\ }\textbf {\bibinfo {volume} {67}},\ \bibinfo {pages} {30} (\bibinfo
  {year} {2014})}\BibitemShut {NoStop}%
\bibitem [{\citenamefont {Giovannetti}\ \emph {et~al.}(2004)\citenamefont
  {Giovannetti}, \citenamefont {Lloyd},\ and\ \citenamefont
  {Maccone}}]{GLM04Quantum}%
  \BibitemOpen
  \bibfield  {author} {\bibinfo {author} {\bibfnamefont {V.}~\bibnamefont
  {Giovannetti}}, \bibinfo {author} {\bibfnamefont {S.}~\bibnamefont {Lloyd}},
  \ and\ \bibinfo {author} {\bibfnamefont {L.}~\bibnamefont {Maccone}},\
  }\href@noop {} {\bibfield  {journal} {\bibinfo  {journal} {Science}\ }\textbf
  {\bibinfo {volume} {306}},\ \bibinfo {pages} {1330} (\bibinfo {year}
  {2004})}\BibitemShut {NoStop}%
\bibitem [{\citenamefont {Castin}\ and\ \citenamefont {Dum}(1996)}]{CD96Bose}%
  \BibitemOpen
  \bibfield  {author} {\bibinfo {author} {\bibfnamefont {Y.}~\bibnamefont
  {Castin}}\ and\ \bibinfo {author} {\bibfnamefont {R.}~\bibnamefont {Dum}},\
  }\href@noop {} {\bibfield  {journal} {\bibinfo  {journal} {Phys. Rev. Lett.}\
  }\textbf {\bibinfo {volume} {77}},\ \bibinfo {pages} {5315} (\bibinfo {year}
  {1996})}\BibitemShut {NoStop}%
\bibitem [{\citenamefont {Weinacht}\ \emph {et~al.}(1999)\citenamefont
  {Weinacht}, \citenamefont {Ahn},\ and\ \citenamefont
  {Bucksbaum}}]{WAB99Controlling}%
  \BibitemOpen
  \bibfield  {author} {\bibinfo {author} {\bibfnamefont {T.~C.}\ \bibnamefont
  {Weinacht}}, \bibinfo {author} {\bibfnamefont {J.}~\bibnamefont {Ahn}}, \
  and\ \bibinfo {author} {\bibfnamefont {P.~H.}\ \bibnamefont {Bucksbaum}},\
  }\href@noop {} {\bibfield  {journal} {\bibinfo  {journal} {Nature}\ }\textbf
  {\bibinfo {volume} {397}},\ \bibinfo {pages} {233} (\bibinfo {year}
  {1999})}\BibitemShut {NoStop}%
\bibitem [{\citenamefont {Olshanii}\ \emph {et~al.}(2000)\citenamefont
  {Olshanii}, \citenamefont {Dekker}, \citenamefont {Herzog},\ and\
  \citenamefont {Prentiss}}]{ODHP00deBroglie}%
  \BibitemOpen
  \bibfield  {author} {\bibinfo {author} {\bibfnamefont {M.}~\bibnamefont
  {Olshanii}}, \bibinfo {author} {\bibfnamefont {N.}~\bibnamefont {Dekker}},
  \bibinfo {author} {\bibfnamefont {C.}~\bibnamefont {Herzog}}, \ and\ \bibinfo
  {author} {\bibfnamefont {M.}~\bibnamefont {Prentiss}},\ }\href@noop {}
  {\bibfield  {journal} {\bibinfo  {journal} {Phys. Rev. A}\ }\textbf {\bibinfo
  {volume} {62}},\ \bibinfo {pages} {033612} (\bibinfo {year}
  {2000})}\BibitemShut {NoStop}%
\bibitem [{\citenamefont {Eiermann}\ \emph {et~al.}(2003)\citenamefont
  {Eiermann}, \citenamefont {Treutlein}, \citenamefont {Anker}, \citenamefont
  {Albiez}, \citenamefont {Taglieber}, \citenamefont {Marzlin}, ,\ and\
  \citenamefont {Oberthaler}}]{ETA+03Dispersion}%
  \BibitemOpen
  \bibfield  {author} {\bibinfo {author} {\bibfnamefont {B.}~\bibnamefont
  {Eiermann}}, \bibinfo {author} {\bibfnamefont {P.}~\bibnamefont {Treutlein}},
  \bibinfo {author} {\bibfnamefont {T.}~\bibnamefont {Anker}}, \bibinfo
  {author} {\bibfnamefont {M.}~\bibnamefont {Albiez}}, \bibinfo {author}
  {\bibfnamefont {M.}~\bibnamefont {Taglieber}}, \bibinfo {author}
  {\bibfnamefont {K.-P.}\ \bibnamefont {Marzlin}}, , \ and\ \bibinfo {author}
  {\bibfnamefont {M.~K.}\ \bibnamefont {Oberthaler}},\ }\href@noop {}
  {\bibfield  {journal} {\bibinfo  {journal} {Phys. Rev. Lett.}\ }\textbf
  {\bibinfo {volume} {91}},\ \bibinfo {pages} {060402} (\bibinfo {year}
  {2003})}\BibitemShut {NoStop}%
\bibitem [{\citenamefont {Nest}\ \emph {et~al.}(2010)\citenamefont {Nest},
  \citenamefont {Japha}, \citenamefont {Folman},\ and\ \citenamefont
  {Kosloff}}]{NJF+10Dynamic}%
  \BibitemOpen
  \bibfield  {author} {\bibinfo {author} {\bibfnamefont {M.}~\bibnamefont
  {Nest}}, \bibinfo {author} {\bibfnamefont {Y.}~\bibnamefont {Japha}},
  \bibinfo {author} {\bibfnamefont {R.}~\bibnamefont {Folman}}, \ and\ \bibinfo
  {author} {\bibfnamefont {R.}~\bibnamefont {Kosloff}},\ }\href@noop {}
  {\bibfield  {journal} {\bibinfo  {journal} {Phys. Rev. A}\ }\textbf {\bibinfo
  {volume} {81}},\ \bibinfo {pages} {043632} (\bibinfo {year}
  {2010})}\BibitemShut {NoStop}%
\bibitem [{\citenamefont {Fabre}\ \emph {et~al.}(2011)\citenamefont {Fabre},
  \citenamefont {Cheiney}, \citenamefont {Gattobigio}, \citenamefont
  {Vermersch}, \citenamefont {Faure}, \citenamefont {Mathevet}, \citenamefont
  {Lahaye}, ,\ and\ \citenamefont {{Gu\'{e}ry-Odelin}}}]{FCG+11Realization}%
  \BibitemOpen
  \bibfield  {author} {\bibinfo {author} {\bibfnamefont {C.~M.}\ \bibnamefont
  {Fabre}}, \bibinfo {author} {\bibfnamefont {P.}~\bibnamefont {Cheiney}},
  \bibinfo {author} {\bibfnamefont {G.~L.}\ \bibnamefont {Gattobigio}},
  \bibinfo {author} {\bibfnamefont {F.}~\bibnamefont {Vermersch}}, \bibinfo
  {author} {\bibfnamefont {S.}~\bibnamefont {Faure}}, \bibinfo {author}
  {\bibfnamefont {R.}~\bibnamefont {Mathevet}}, \bibinfo {author}
  {\bibfnamefont {T.}~\bibnamefont {Lahaye}}, , \ and\ \bibinfo {author}
  {\bibfnamefont {D.}~\bibnamefont {{Gu\'{e}ry-Odelin}}},\ }\href@noop {}
  {\bibfield  {journal} {\bibinfo  {journal} {Phys. Rev. Lett.}\ }\textbf
  {\bibinfo {volume} {107}},\ \bibinfo {pages} {230401} (\bibinfo {year}
  {2011})}\BibitemShut {NoStop}%
\bibitem [{\citenamefont {Cheiney}\ \emph {et~al.}(2013)\citenamefont
  {Cheiney}, \citenamefont {Fabre}, \citenamefont {Vermersch}, \citenamefont
  {Gattobigio}, \citenamefont {Mathevet}, \citenamefont {Lahaye},\ and\
  \citenamefont {{Gu\'{e}ry-Odelin}}}]{CFV+13Matter}%
  \BibitemOpen
  \bibfield  {author} {\bibinfo {author} {\bibfnamefont {P.}~\bibnamefont
  {Cheiney}}, \bibinfo {author} {\bibfnamefont {C.~M.}\ \bibnamefont {Fabre}},
  \bibinfo {author} {\bibfnamefont {F.}~\bibnamefont {Vermersch}}, \bibinfo
  {author} {\bibfnamefont {G.~L.}\ \bibnamefont {Gattobigio}}, \bibinfo
  {author} {\bibfnamefont {R.}~\bibnamefont {Mathevet}}, \bibinfo {author}
  {\bibfnamefont {T.}~\bibnamefont {Lahaye}}, \ and\ \bibinfo {author}
  {\bibfnamefont {D.}~\bibnamefont {{Gu\'{e}ry-Odelin}}},\ }\href@noop {}
  {\bibfield  {journal} {\bibinfo  {journal} {Phys. Rev. A}\ }\textbf {\bibinfo
  {volume} {87}},\ \bibinfo {pages} {013623} (\bibinfo {year}
  {2013})}\BibitemShut {NoStop}%
\bibitem [{\citenamefont {Gattobigio}\ \emph {et~al.}(2012)\citenamefont
  {Gattobigio}, \citenamefont {Couvert}, \citenamefont {Reinaudi},
  \citenamefont {Georgeot},\ and\ \citenamefont
  {{Gu\'{e}ry-Odelin}}}]{GCR+12Optically}%
  \BibitemOpen
  \bibfield  {author} {\bibinfo {author} {\bibfnamefont {G.~L.}\ \bibnamefont
  {Gattobigio}}, \bibinfo {author} {\bibfnamefont {A.}~\bibnamefont {Couvert}},
  \bibinfo {author} {\bibfnamefont {G.}~\bibnamefont {Reinaudi}}, \bibinfo
  {author} {\bibfnamefont {B.}~\bibnamefont {Georgeot}}, \ and\ \bibinfo
  {author} {\bibfnamefont {D.}~\bibnamefont {{Gu\'{e}ry-Odelin}}},\ }\href@noop
  {} {\bibfield  {journal} {\bibinfo  {journal} {Phys. Rev. Lett.}\ }\textbf
  {\bibinfo {volume} {109}},\ \bibinfo {pages} {030403} (\bibinfo {year}
  {2012})}\BibitemShut {NoStop}%
\bibitem [{\citenamefont {Muga}\ \emph {et~al.}(2004)\citenamefont {Muga},
  \citenamefont {Palao}, \citenamefont {Navarro},\ and\ \citenamefont
  {Egusquiza}}]{MPNE04Complex}%
  \BibitemOpen
  \bibfield  {author} {\bibinfo {author} {\bibfnamefont {J.~G.}\ \bibnamefont
  {Muga}}, \bibinfo {author} {\bibfnamefont {J.~P.}\ \bibnamefont {Palao}},
  \bibinfo {author} {\bibfnamefont {B.}~\bibnamefont {Navarro}}, \ and\
  \bibinfo {author} {\bibfnamefont {I.~L.}\ \bibnamefont {Egusquiza}},\
  }\href@noop {} {\bibfield  {journal} {\bibinfo  {journal} {Phys. Rep.}\
  }\textbf {\bibinfo {volume} {395}},\ \bibinfo {pages} {357} (\bibinfo {year}
  {2004})}\BibitemShut {NoStop}%
\bibitem [{\citenamefont {Kleber}(1994)}]{Kle94Exact}%
  \BibitemOpen
  \bibfield  {author} {\bibinfo {author} {\bibfnamefont {M.}~\bibnamefont
  {Kleber}},\ }\href@noop {} {\bibfield  {journal} {\bibinfo  {journal} {Phys.
  Rep.}\ }\textbf {\bibinfo {volume} {236}},\ \bibinfo {pages} {331} (\bibinfo
  {year} {1994})}\BibitemShut {NoStop}%
\bibitem [{\citenamefont {{del Campo}}\ \emph {et~al.}(2009)\citenamefont {{del
  Campo}}, \citenamefont {Garc{\'i}a-{C}alder{\'o}n},\ and\ \citenamefont
  {Muga}}]{CGM09Quantum}%
  \BibitemOpen
  \bibfield  {author} {\bibinfo {author} {\bibfnamefont {A.}~\bibnamefont {{del
  Campo}}}, \bibinfo {author} {\bibfnamefont {G.}~\bibnamefont
  {Garc{\'i}a-{C}alder{\'o}n}}, \ and\ \bibinfo {author} {\bibfnamefont
  {J.~G.}\ \bibnamefont {Muga}},\ }\href@noop {} {\bibfield  {journal}
  {\bibinfo  {journal} {Phys. Rep.}\ }\textbf {\bibinfo {volume} {476}},\
  \bibinfo {pages} {1} (\bibinfo {year} {2009})}\BibitemShut {NoStop}%
\bibitem [{\citenamefont {Moshinsky}(1952)}]{Mos52Diffraction}%
  \BibitemOpen
  \bibfield  {author} {\bibinfo {author} {\bibfnamefont {M.}~\bibnamefont
  {Moshinsky}},\ }\href@noop {} {\bibfield  {journal} {\bibinfo  {journal}
  {Phys. Rev.}\ }\textbf {\bibinfo {volume} {88}},\ \bibinfo {pages} {625}
  (\bibinfo {year} {1952})}\BibitemShut {NoStop}%
\bibitem [{\citenamefont {Gerasimov}\ and\ \citenamefont
  {Kazarnovskii}(1976)}]{GK76Possibility}%
  \BibitemOpen
  \bibfield  {author} {\bibinfo {author} {\bibfnamefont {A.~S.}\ \bibnamefont
  {Gerasimov}}\ and\ \bibinfo {author} {\bibfnamefont {M.~V.}\ \bibnamefont
  {Kazarnovskii}},\ }\href@noop {} {\bibfield  {journal} {\bibinfo  {journal}
  {Sov. Phys. JETP}\ }\textbf {\bibinfo {volume} {44}},\ \bibinfo {pages} {892}
  (\bibinfo {year} {1976})}\BibitemShut {NoStop}%
\bibitem [{\citenamefont {Moshinsky}(1976)}]{Mos76Diffraction}%
  \BibitemOpen
  \bibfield  {author} {\bibinfo {author} {\bibfnamefont {M.}~\bibnamefont
  {Moshinsky}},\ }\href@noop {} {\bibfield  {journal} {\bibinfo  {journal} {Am.
  J. Phys.}\ }\textbf {\bibinfo {volume} {44}},\ \bibinfo {pages} {1037}
  (\bibinfo {year} {1976})}\BibitemShut {NoStop}%
\bibitem [{\citenamefont {Brukner}\ and\ \citenamefont
  {Zeilinger}(1997)}]{BZ97Diffraction}%
  \BibitemOpen
  \bibfield  {author} {\bibinfo {author} {\bibfnamefont {C.}~\bibnamefont
  {Brukner}}\ and\ \bibinfo {author} {\bibfnamefont {A.}~\bibnamefont
  {Zeilinger}},\ }\href@noop {} {\bibfield  {journal} {\bibinfo  {journal}
  {Phys. Rev. A}\ }\textbf {\bibinfo {volume} {56}},\ \bibinfo {pages} {3804}
  (\bibinfo {year} {1997})}\BibitemShut {NoStop}%
\bibitem [{\citenamefont {Man'ko}\ \emph {et~al.}(1999)\citenamefont {Man'ko},
  \citenamefont {Moshinsky},\ and\ \citenamefont {Sharma}}]{MMS99Diffraction}%
  \BibitemOpen
  \bibfield  {author} {\bibinfo {author} {\bibfnamefont {V.}~\bibnamefont
  {Man'ko}}, \bibinfo {author} {\bibfnamefont {M.}~\bibnamefont {Moshinsky}}, \
  and\ \bibinfo {author} {\bibfnamefont {A.}~\bibnamefont {Sharma}},\
  }\href@noop {} {\bibfield  {journal} {\bibinfo  {journal} {Phys. Rev. A}\
  }\textbf {\bibinfo {volume} {59}},\ \bibinfo {pages} {1809} (\bibinfo {year}
  {1999})}\BibitemShut {NoStop}%
\bibitem [{\citenamefont {Godoy}(2002)}]{God02Diffraction}%
  \BibitemOpen
  \bibfield  {author} {\bibinfo {author} {\bibfnamefont {S.}~\bibnamefont
  {Godoy}},\ }\href@noop {} {\bibfield  {journal} {\bibinfo  {journal} {Phys.
  Rev. A}\ }\textbf {\bibinfo {volume} {65}},\ \bibinfo {pages} {042111}
  (\bibinfo {year} {2002})}\BibitemShut {NoStop}%
\bibitem [{\citenamefont {Godoy}(2003)}]{God03Diffraction}%
  \BibitemOpen
  \bibfield  {author} {\bibinfo {author} {\bibfnamefont {S.}~\bibnamefont
  {Godoy}},\ }\href@noop {} {\bibfield  {journal} {\bibinfo  {journal} {Phys.
  Rev. A}\ }\textbf {\bibinfo {volume} {67}},\ \bibinfo {pages} {012102}
  (\bibinfo {year} {2003})}\BibitemShut {NoStop}%
\bibitem [{\citenamefont {Granot}\ and\ \citenamefont
  {Marchewka}(2005)}]{GM05Generic}%
  \BibitemOpen
  \bibfield  {author} {\bibinfo {author} {\bibfnamefont {E.}~\bibnamefont
  {Granot}}\ and\ \bibinfo {author} {\bibfnamefont {A.}~\bibnamefont
  {Marchewka}},\ }\href@noop {} {\bibfield  {journal} {\bibinfo  {journal} {EPL
  (Europhys. Lett.)}\ }\textbf {\bibinfo {volume} {72}},\ \bibinfo {pages}
  {341} (\bibinfo {year} {2005})}\BibitemShut {NoStop}%
\bibitem [{\citenamefont {Torrontegui}\ \emph {et~al.}(2011)\citenamefont
  {Torrontegui}, \citenamefont {Mu{\~n}oz}, \citenamefont {Ban},\ and\
  \citenamefont {Muga}}]{TMB+11Explanation}%
  \BibitemOpen
  \bibfield  {author} {\bibinfo {author} {\bibfnamefont {E.}~\bibnamefont
  {Torrontegui}}, \bibinfo {author} {\bibfnamefont {J.}~\bibnamefont
  {Mu{\~n}oz}}, \bibinfo {author} {\bibfnamefont {Y.}~\bibnamefont {Ban}}, \
  and\ \bibinfo {author} {\bibfnamefont {J.~G.}\ \bibnamefont {Muga}},\
  }\href@noop {} {\bibfield  {journal} {\bibinfo  {journal} {Phys. Rev. A}\
  }\textbf {\bibinfo {volume} {83}},\ \bibinfo {pages} {043608} (\bibinfo
  {year} {2011})}\BibitemShut {NoStop}%
\bibitem [{\citenamefont {{del Campo}}\ \emph {et~al.}(2007)\citenamefont {{del
  Campo}}, \citenamefont {Muga},\ and\ \citenamefont {Moshinsky}}]{CMM07Time}%
  \BibitemOpen
  \bibfield  {author} {\bibinfo {author} {\bibfnamefont {A.}~\bibnamefont {{del
  Campo}}}, \bibinfo {author} {\bibfnamefont {J.~G.}\ \bibnamefont {Muga}}, \
  and\ \bibinfo {author} {\bibfnamefont {M.}~\bibnamefont {Moshinsky}},\
  }\href@noop {} {\bibfield  {journal} {\bibinfo  {journal} {J. Phys. B: At.
  Mol. Opt. Phys.}\ }\textbf {\bibinfo {volume} {40}},\ \bibinfo {pages} {975}
  (\bibinfo {year} {2007})}\BibitemShut {NoStop}%
\bibitem [{\citenamefont {Godoy}(2009)}]{God09Statistical}%
  \BibitemOpen
  \bibfield  {author} {\bibinfo {author} {\bibfnamefont {S.}~\bibnamefont
  {Godoy}},\ }\href@noop {} {\bibfield  {journal} {\bibinfo  {journal} {Physica
  B}\ }\textbf {\bibinfo {volume} {404}},\ \bibinfo {pages} {1826} (\bibinfo
  {year} {2009})}\BibitemShut {NoStop}%
\bibitem [{\citenamefont {{del Campo}}\ and\ \citenamefont
  {Muga}(2006)}]{CM06Dynamics}%
  \BibitemOpen
  \bibfield  {author} {\bibinfo {author} {\bibfnamefont {A.}~\bibnamefont {{del
  Campo}}}\ and\ \bibinfo {author} {\bibfnamefont {J.~G.}\ \bibnamefont
  {Muga}},\ }\href@noop {} {\bibfield  {journal} {\bibinfo  {journal} {EPL
  (Europhys. Lett.)}\ }\textbf {\bibinfo {volume} {74}},\ \bibinfo {pages}
  {965} (\bibinfo {year} {2006})}\BibitemShut {NoStop}%
\bibitem [{\citenamefont {Goussev}(2012)}]{Gou12Huygens}%
  \BibitemOpen
  \bibfield  {author} {\bibinfo {author} {\bibfnamefont {A.}~\bibnamefont
  {Goussev}},\ }\href@noop {} {\bibfield  {journal} {\bibinfo  {journal} {Phys.
  Rev. A}\ }\textbf {\bibinfo {volume} {85}},\ \bibinfo {pages} {013626}
  (\bibinfo {year} {2012})}\BibitemShut {NoStop}%
\bibitem [{\citenamefont {Goussev}(2013)}]{Gou13Diffraction}%
  \BibitemOpen
  \bibfield  {author} {\bibinfo {author} {\bibfnamefont {A.}~\bibnamefont
  {Goussev}},\ }\href@noop {} {\bibfield  {journal} {\bibinfo  {journal} {Phys.
  Rev. A}\ }\textbf {\bibinfo {volume} {87}},\ \bibinfo {pages} {053621}
  (\bibinfo {year} {2013})}\BibitemShut {NoStop}%
\bibitem [{\citenamefont {Kottler}(1923)}]{Kot23Zur}%
  \BibitemOpen
  \bibfield  {author} {\bibinfo {author} {\bibfnamefont {F.}~\bibnamefont
  {Kottler}},\ }\href@noop {} {\bibfield  {journal} {\bibinfo  {journal} {Ann.
  Phys. (Leipzig)}\ }\textbf {\bibinfo {volume} {70}},\ \bibinfo {pages} {405} (\bibinfo
  {year} {1923})}\BibitemShut {NoStop}%
\bibitem [{\citenamefont {Kottler}(1965)}]{Kot65Diffraction}%
  \BibitemOpen
  \bibfield  {author} {\bibinfo {author} {\bibfnamefont {F.}~\bibnamefont
  {Kottler}},\ }\href@noop {} {\bibfield  {journal} {\bibinfo  {journal} {Prog.
  Opt.}\ }\textbf {\bibinfo {volume} {4}},\ \bibinfo {pages} {281} (\bibinfo
  {year} {1965})}\BibitemShut {NoStop}%
\bibitem [{\citenamefont {Ballentine}(2014)}]{Bal14Quantum}%
  \BibitemOpen
  \bibfield  {author} {\bibinfo {author} {\bibfnamefont {L.}~\bibnamefont
  {Ballentine}},\ }\href@noop {} {\emph {\bibinfo {title} {Quantum Mechanics: A
  Modern Development}}},\ \bibinfo {edition} {2nd}\ ed.\ (\bibinfo  {publisher}
  {World Scientific, Singapore},\ \bibinfo {year} {2014})\BibitemShut {NoStop}%
\bibitem [{\citenamefont {Heller}(1981)}]{Hel81Frozen}%
  \BibitemOpen
  \bibfield  {author} {\bibinfo {author} {\bibfnamefont {E.~J.}\ \bibnamefont
  {Heller}},\ }\href@noop {} {\bibfield  {journal} {\bibinfo  {journal} {J.
  Chem. Phys.}\ }\textbf {\bibinfo {volume} {75}},\ \bibinfo {pages} {2923}
  (\bibinfo {year} {1981})}\BibitemShut {NoStop}%
\bibitem [{\citenamefont {Jendrzejewski}\ \emph {et~al.}(2012)\citenamefont
  {Jendrzejewski}, \citenamefont {M\"{u}ller}, \citenamefont {Richard},
  \citenamefont {Date}, \citenamefont {Plisson}, \citenamefont {Bouyer},
  \citenamefont {Aspect},\ and\ \citenamefont {Josse}}]{JMR+12Coherent}%
  \BibitemOpen
  \bibfield  {author} {\bibinfo {author} {\bibfnamefont {F.}~\bibnamefont
  {Jendrzejewski}}, \bibinfo {author} {\bibfnamefont {K.}~\bibnamefont
  {M\"{u}ller}}, \bibinfo {author} {\bibfnamefont {J.}~\bibnamefont {Richard}},
  \bibinfo {author} {\bibfnamefont {A.}~\bibnamefont {Date}}, \bibinfo {author}
  {\bibfnamefont {T.}~\bibnamefont {Plisson}}, \bibinfo {author} {\bibfnamefont
  {P.}~\bibnamefont {Bouyer}}, \bibinfo {author} {\bibfnamefont
  {A.}~\bibnamefont {Aspect}}, \ and\ \bibinfo {author} {\bibfnamefont
  {V.}~\bibnamefont {Josse}},\ }\href@noop {} {\bibfield  {journal} {\bibinfo
  {journal} {Phys. Rev. Lett.}\ }\textbf {\bibinfo {volume} {109}},\ \bibinfo
  {pages} {195302} (\bibinfo {year} {2012})}\BibitemShut {NoStop}%
\bibitem [{\citenamefont {Anderson}\ \emph {et~al.}(1995)\citenamefont
  {Anderson}, \citenamefont {Ensher}, \citenamefont {Matthews}, \citenamefont
  {Wieman},\ and\ \citenamefont {Cornell}}]{AEM+95Observation}%
  \BibitemOpen
  \bibfield  {author} {\bibinfo {author} {\bibfnamefont {M.~H.}\ \bibnamefont
  {Anderson}}, \bibinfo {author} {\bibfnamefont {J.~R.}\ \bibnamefont
  {Ensher}}, \bibinfo {author} {\bibfnamefont {M.~R.}\ \bibnamefont
  {Matthews}}, \bibinfo {author} {\bibfnamefont {C.~E.}\ \bibnamefont
  {Wieman}}, \ and\ \bibinfo {author} {\bibfnamefont {E.~A.}\ \bibnamefont
  {Cornell}},\ }\href@noop {} {\bibfield  {journal} {\bibinfo  {journal}
  {Science}\ }\textbf {\bibinfo {volume} {269}},\ \bibinfo {pages} {198}
  (\bibinfo {year} {1995})}\BibitemShut {NoStop}%
\end{thebibliography}
\end{document}